\documentclass[12pt]{article}

\usepackage{epsfig}

\newcommand{\mm}{M}      
\newcommand{\ee}{m_\ell} 
\newcommand{\qq}{Q'}     

\newcommand{\re}{\mathrm{Re}\,}
\newcommand{\im}{\mathrm{Im}\,}

\newcommand{\gev}{\,{\rm GeV}}

\newcommand{\dd}{D\!D}   

\begin{document}

\begin{flushright}
  CPHT--S010--0201 \\
  DESY--01--119 \\
  hep-ph/0110062
\end{flushright}

\vspace{\baselineskip}

\begin{center}
\textbf{\LARGE Timelike Compton scattering: \\[0.5\baselineskip]
exclusive photoproduction of lepton pairs
} \\
\vspace{3\baselineskip}
{\large
E. R. Berger\,${}^{a,}$\footnote{Email:
edgar.berger@cpht.polytechnique.fr}, 
M. Diehl\,${}^{b,}$\footnote{Email: markus.diehl@desy.de},
B. Pire\,${}^{a,}$\footnote{Email: bernard.pire@cpht.polytechnique.fr}}
\\
\vspace{2\baselineskip}
${}^a$\,CPHT\footnote{Unit{\'e} mixte C7644 du CNRS}, {\'E}cole
Polytechnique, 91128 Palaiseau, France \\[0.5\baselineskip]
${}^b$\,Deutsches Elektronen-Synchroton DESY, 22603 Hamburg, Germany
\\
\textit{present address:} \\
Institut f\"ur Theoretische Physik E, RWTH Aachen, 52056 Aachen,
Germany \\
\vspace{5\baselineskip}
\textbf{Abstract}\\
\vspace{1\baselineskip}
\parbox{0.9\textwidth}{We investigate the exclusive photoproduction of
a heavy timelike photon which decays into a lepton pair, $\gamma p\to
\ell^+\!\ell^- \,p$.  This can be seen as the analog of deeply virtual
Compton scattering, and we argue that the two processes are
complementary for studying generalized parton distributions in the
nucleon.  In an unpolarized experiment the angular distribution of the
leptons readily provides access to the real part of the Compton
amplitude.  We estimate the possible size of this effect in kinematics
where the Compton process should be dominated by quark exchange.}
\end{center}

\newpage

\section{Introduction}

A considerable amount of theoretical and experimental work is
currently being devoted to the study of generalized parton
distributions, whose measurement could make important contributions to
our understanding of how quarks and gluons assemble themselves to
hadrons~\cite{Muller:1994fv,Ji:1997ek,Radyushkin:1997ki}. The
theoretically simplest and cleanest of the exclusive processes where
these distributions occur is deeply virtual Compton scattering (DVCS),
i.e., $\gamma^* p \to \gamma p$ in kinematics where the $\gamma^*$ has
large spacelike virtuality while the invariant momentum transfer $t$
to the proton is small. In the present paper, we investigate the
``inverse'' process, $\gamma p \to \gamma^* p$ at small $t$ and large
\emph{timelike} virtuality of the final state photon. We shall refer
to this as timelike Compton scattering (TCS). This reaction shares
many features of DVCS, although the timelike character of the virtual
photon entails some specific differences. The combination of data on
DVCS and TCS would offer a powerful tool to make sure we understand
the reaction mechanism, and eventually to obtain stronger constraints
on the generalized parton distributions than DVCS alone would provide.

%
\begin{figure}[htb]
\begin{center}
     \epsfxsize=0.5\textwidth
     \epsffile{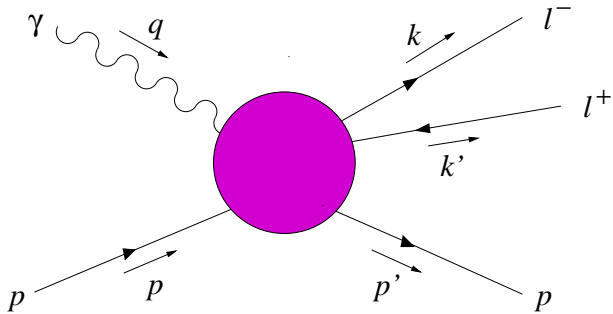}
\caption{\label{refig}Real photon-proton scattering into a lepton pair
and a proton. $\ell$ stands for an electron or a muon.}
\end{center}
\end{figure}
%

The physical process where to observe TCS is photoproduction of a
heavy lepton pair, $\gamma p \to \mu^+\!\mu^-\, p$ or $\gamma p \to
e^+\!e^-\, p$, shown in Fig.~\ref{refig}. Despite the close analogy to
real photon production $e p \to e \gamma p$ or $\mu p \to \mu \gamma
p$, where DVCS can be accessed, the phenomenology of these reactions
shows important differences. In both cases, a Bethe-Heitler (BH)
mechanism contributes at the amplitude level. Contrary to the case of
DVCS, this contribution \emph{always} dominates over the one from TCS
in the kinematical regime where we want to study it. On the other
hand, the interference between the TCS and BH processes can readily be
accessed through the angular distribution of the lepton pair, whereas
the corresponding observable for DVCS is the lepton charge asymmetry
and requires beams of both positive and negative charge.

This paper is organized as follows.  In section \ref{sec:compton} we
review the kinematics, factorization properties, and helicity
structure of the Compton amplitude in the general case where the two
photon virtualities are different, but at least one of them is
sufficiently large to provide a hard scale.  In section \ref{sec:time}
we discuss specific features related to the timelike nature of the
outgoing photon in TCS.  We develop the phenomenology of exclusive
photoproduction of a lepton pair in section \ref{sec:pair}, taking
into account the Bethe-Heitler and the Compton processes and their
interference.  In section \ref{sec:numbers} we present estimates of
cross sections and of asymmetries suitable to extract information on
the Compton signal.  Section \ref{sec:sum} contains our conclusions.
In an appendix we discuss the relevance of parton densities at very
small $x$ when modeling generalized parton distributions with a double
distribution ansatz.

\section{The Compton amplitude}
\label{sec:compton}

\begin{figure}
\begin{center}
    \epsfxsize=0.39\textwidth \epsffile{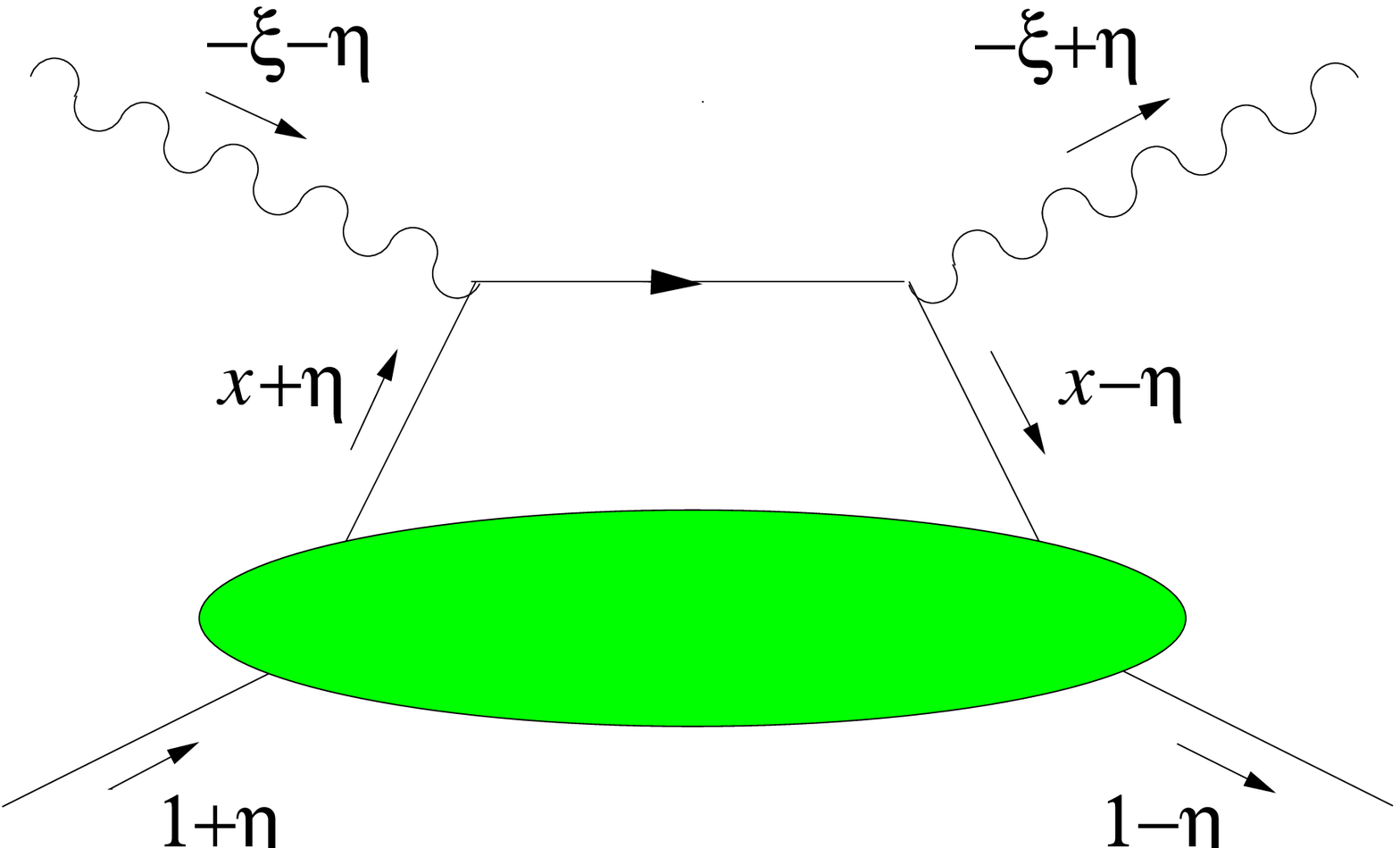}
\hspace{0.05\textwidth}
    \epsfxsize=0.39\textwidth
    \epsffile{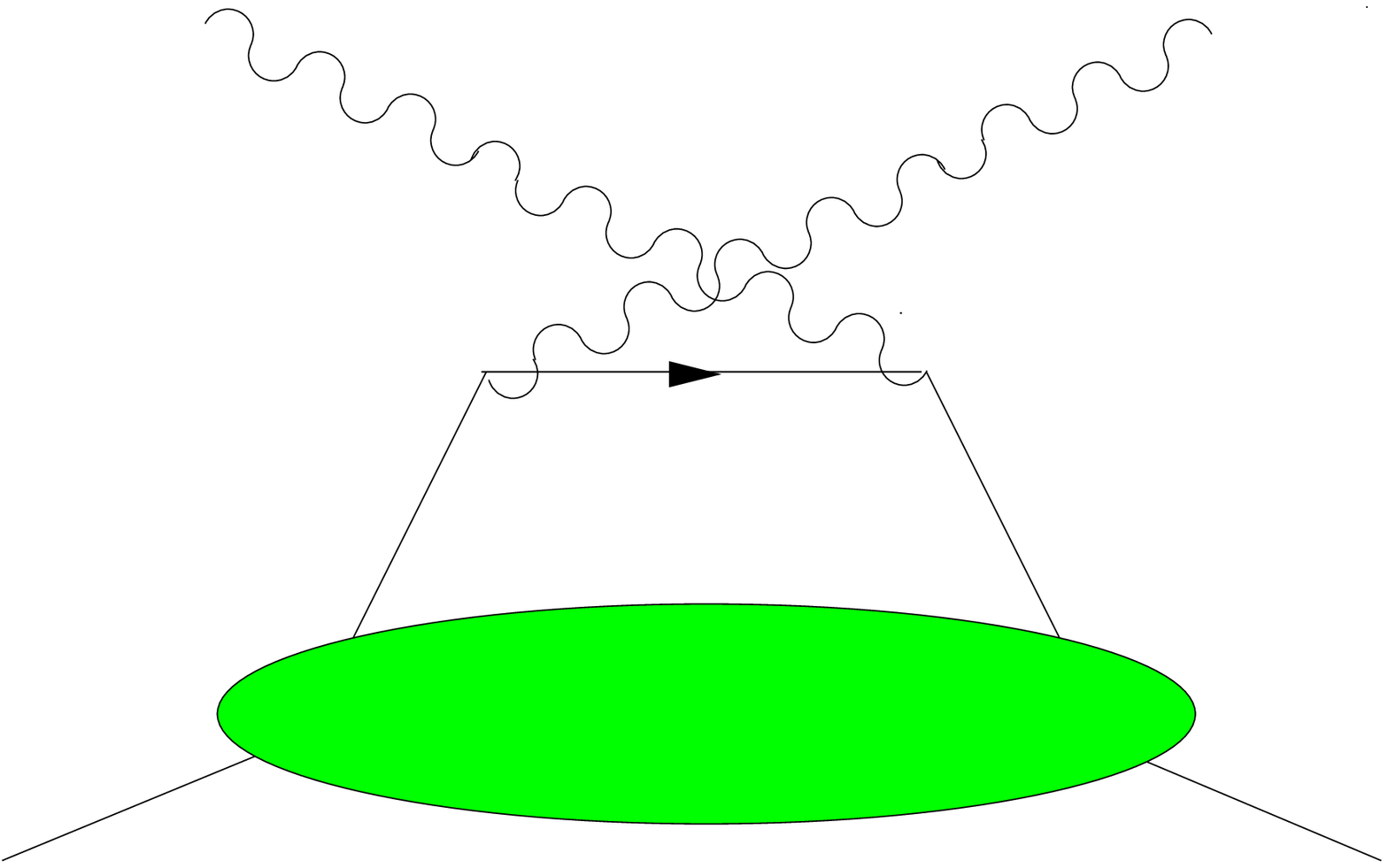}
\caption{Handbag diagrams for the Compton process
(\protect\ref{general-compton}) in the scaling limit. The
plus-momentum fractions $x$, $\xi$, $\eta$ refer to the average proton
momentum $\frac{1}{2}(p+p')$.}
\label{haba}
\end{center}
\end{figure}

Both DVCS and TCS are limiting cases of the general Compton process
\begin{equation}
  \gamma^*(q) + p(p) \to \gamma^*(q') + p(p') ,
\label{general-compton}
\end{equation}
where the four-momenta $q$ and $q'$ of the photons can have any
virtuality.  We will also use $\Delta = p' - p$, the invariants
\begin{equation}
Q^2 = - q^2 , \qquad Q'^2 = q'^2 , \qquad
s = (p+q)^2 , \qquad t = \Delta^2 ,
\label{basic-kin}
\end{equation}
and write $M$ for the proton mass.  In the region where at least one
of the virtualities is large, the amplitude is given by the
convolution of hard scattering coefficients, calculable in
perturbation theory, and generalized parton distributions, which
describe the nonperturbative physics of the process. To leading order
in $\alpha_s$ one then has the quark handbag diagrams of
Fig.~\ref{haba}.  The arguments for factorization given
in~\cite{Collins:1999be}, based on the analysis of Feynman graphs,
hold both for large spacelike and for large timelike
virtualities~\cite{Collins:2001}.\footnote{In contrast, approaches
based on the operator production expansion
\protect\cite{Muller:1994fv} require $(q+q')^2$ to be spacelike, which
is equivalent to $q^2 + q'^2 < 0$ when $t$ can be neglected.}
We thus define the scaling limit as $|q^2| + |q'^2| \to \infty$ at
fixed $t$ and fixed ratios $q^2 /s$ and $q'^2/s$.

%
%

For our subsequent discussion let us recall the expression of the
hadronic tensor
\begin{equation}
T^{\alpha\beta} = i \int d^4x\, e^{-iq\cdot x} \langle p(p')|\, T
J^\alpha_{em}(x) J^\beta_{em}(0) \,| p(p)\rangle ,
\end{equation}
where $e J^\alpha_{em}(x)$ is the electromagnetic current with $e$
denoting the positron charge. In the scaling limit we have to leading
order in $\alpha_s$
\begin{equation}
T^{\alpha\beta} = - \frac{1}{(p+p')^+}\, \bar{u}(p^{\prime}) 
\left[\,
   g_T^{\alpha\beta} \, \Big(
      {\cal{H}}_1 \, \gamma^+ +
      {\cal{E}}_1 \, \frac{i \sigma^{+\rho}\Delta_{\rho}}{2 \mm}
   \Big)
   + i \epsilon_T^{\alpha\beta} \, \Big(
      \tilde{\cal{H}}_1 \, \gamma^+ \gamma_5 +
      \tilde{\cal{E}}_1 \, \frac{\Delta^+ \gamma_5}{2 \mm}
   \Big) 
\,\right] u(p) .
\label{leading-order}
\end{equation}
This expression holds in reference frames where both proton momenta
$p$ and $p'$ have small transverse components of order $\sqrt{-t}$ and
are moving fast to the right, i.e., have large plus-components.
Light-cone coordinates are defined as $v^\pm = (v^0\pm v^3)/ \sqrt{2}$
for any four-vector~$v$. The transverse tensors $g_T$ and $\epsilon_T$
have as only nonzero components $-g_T^{11} = -g_T^{22} =
\epsilon_T^{12} = - \epsilon_T^{21} = 1$. Following the notation of
\cite{Belitsky:2001gz} we have introduced the convolutions
\begin{eqnarray}
 {\cal H}_1(\xi,\eta,t) &=& \sum_q e_q^2 \int_{-1}^{1} d x
    \Big( \frac{1}{\xi-x-i\epsilon} - \frac{1}{\xi+x-i\epsilon} \Big)
    H^q(x,\eta,t),
 \nonumber \\
 {\cal E}_1(\xi,\eta,t) &=& \sum_q  e_q^2\int_{-1}^{1}d x
    \Big( \frac{1}{\xi-x-i\epsilon} - \frac{1}{\xi+x-i\epsilon} \Big)
    E^q(x,\eta,t),
 \nonumber \\
 \tilde{\cal H}_1(\xi,\eta,t) &=& \sum_q e_q^2 \int_{-1}^{1}d x
    \Big( \frac{1}{\xi-x-i\epsilon} + \frac{1}{\xi+x-i\epsilon} \Big)
    \tilde H^q(x,\eta,t),
 \nonumber \\
 \tilde{\cal E}_1(\xi,\eta,t) &=& \sum_q e_q^2 \int_{-1}^{1}d x
    \Big( \frac{1}{\xi-x-i\epsilon} + \frac{1}{\xi+x-i\epsilon} \Big)
    \tilde E^q(x,\eta,t),
 \label{htilde}
\end{eqnarray}
of the generalized quark distributions defined in \cite{Ji:1997ek},
summed over quarks of flavor $q$ and electric charge $e e_q$. The
scaling variables $\xi$ and $\eta$ are given by
\begin{eqnarray}
\xi  &=& - \frac{(q+q')^2}{2(p+p')\cdot (q+q')} \,\approx\,
           \frac{Q^2 - Q'^2}{2s + Q^2 - Q'^2} , \nonumber \\
\eta &=& - \frac{(q-q')\cdot (q+q')}{(p+p')\cdot (q+q')} \,\approx\,
           \frac{Q^2 + Q'^2}{2s + Q^2 - Q'^2} ,
\label{xi-eta-def}
\end{eqnarray}
where the approximations hold in the kinematical limit we are working
in. $x$, $\xi$, and $\eta$ represent plus-momentum fractions
\begin{equation}
x = \frac{(k+k')^+}{(p+p')^+} , \qquad
\xi \approx - \frac{(q+q')^+}{(p+p')^+} , \qquad
\eta \approx  \frac{(p-p')^+}{(p+p')^+} .
\end{equation}
The expressions (\ref{leading-order}) and (\ref{htilde}) reveal that
the two-photon amplitude is independent of the photon virtualities at
fixed $\xi$, $\eta$ and $t$.  In the case of spacelike $q = q'$ this
is just Bjorken scaling. To be precise, the independence on $q^2$ and
$q'^2$ only holds up to logarithmic corrections: the photon
virtualities provide the hard scale of the process and thus enter
through the factorization scale dependence of the parton
distributions, which we have not displayed above. The corresponding
evolution equations are well-known
\cite{Muller:1994fv,Ji:1997ek,Radyushkin:1997ki,Belitsky:2000hf}, and
as usual we will refer to $-1<x<-\eta$ and $\eta<x<1$ as the DGLAP
regions, and to $-\eta<x<\eta$ as the ERBL region of the parton
distributions.

Let us now recall the helicity structure of the two-photon process in
the scaling limit. Contracting the hadronic tensor with polarization
vectors $\epsilon$ of the incoming and $\epsilon'$ of the outgoing
photon, one obtains the helicity amplitudes of (\ref{general-compton})
as
\begin{equation}
e^2 M^{\lambda'\mu', \lambda\mu} =
e^2\, \epsilon_\alpha\, T^{\alpha\beta}\, \epsilon'^*_\beta ,
\end{equation}
where $\lambda$ ($\lambda'$) denotes the helicity of the incoming
(outgoing) proton and $\mu$ ($\mu'$) the helicity of the incoming
(outgoing) photon.  Parity invariance provides the relations
$M^{-\lambda'-\mu',-\lambda-\mu} = (-1)^{\lambda'-\mu'-\lambda+\mu}\,
M^{\lambda'\mu', \lambda\mu}$. {}From (\ref{leading-order}) one easily
finds that the quark handbag diagrams only generate helicity
conserving transitions between transverse photons, $M^{\lambda'+,
\lambda+}$ and $M^{\lambda'-, \lambda-}$. At order $\alpha_s$ one
further has amplitudes $M^{\lambda'0, \lambda0}$, provided of course
that both photons are off-shell \cite{Mankiewicz:1998bk}.  Double
helicity flip amplitudes $M^{\lambda'+, \lambda-}$ and $M^{\lambda'-,
\lambda+}$ are generated at order $\alpha_s$ by gluon transversity
distributions \cite{Diehl:1997bu,Hoodbhoy:1998vm}. Finally,
transitions involving one transverse and one longitudinal photon are
suppressed by one power of the large scale $Q$ or $Q'$.  These
twist-three contributions\footnote{We use here the dynamical
definition of twist, where twist $n$ contributions to the Compton
amplitude are suppressed by $n-2$ inverse powers of the large scale.}
have been studied in \cite{Anikin:2000em}, and twist-four
contributions to the double helicity flip amplitudes in
\cite{Kivel:2001rw}. These studies were performed for large spacelike
virtualities; whether they can be extended to the timelike case is a
question beyond the scope of this paper.

The DVCS and TCS processes are limiting cases of
(\ref{general-compton}) where one of the photons is on shell. {}From
(\ref{xi-eta-def}) we readily see that to leading-twist accuracy one
has $\xi = \eta$ in DVCS and $\xi = - \eta$ in TCS. The convolutions
(\ref{htilde}) obey
\begin{eqnarray}
{\cal H}_1(-\eta,\eta,t) = \Big[ {\cal H}_1(\eta,\eta,t) \Big]^* , &&
\tilde{\cal H}_1(-\eta,\eta,t) =
       - \Big[ \tilde{\cal H}_1(\eta,\eta,t) \Big]^* ,
\nonumber \\
{\cal E}_1(-\eta,\eta,t) = \Big[ {\cal E}_1(\eta,\eta,t) \Big]^* , &&
\tilde{\cal E}_1(-\eta,\eta,t) =
       - \Big[ \tilde{\cal E}_1(\eta,\eta,t) \Big]^* ,
\label{TCS-DVCS-conv}
\end{eqnarray}
which leads to the simple relations
\begin{equation}
M^{\lambda'+,\lambda+} \Big|_{TCS}
  = \Big[ M^{\lambda'-,\lambda-} \Big]_{DVCS}^* , \hspace{3em}
M^{\lambda'-,\lambda-} \Big|_{TCS}
  = \Big[ M^{\lambda'+,\lambda+} \Big]_{DVCS}^*
\label{TCS-DVCS-amp}
\end{equation}
between the helicity amplitudes for TCS and DVCS at equal values of
$\eta$ and $t$.  These relations should be evaluated at corresponding
values of $Q'^2$ and $Q^2$ since the photon virtualities play
analogous roles in providing the hard scale of the respective
processes and thus enter in the scale dependence of the parton
distributions.  The relations (\ref{TCS-DVCS-amp}) tell us that at
Born level and to leading twist one obtains the amplitudes for TCS
from those of DVCS by changing the sign of the imaginary part and
reversing the photon polarizations.  To this accuracy, the two
processes thus carry exactly the same information on the generalized
quark distributions.

We remark that the relations (\ref{TCS-DVCS-conv}) and hence
(\ref{TCS-DVCS-amp}) no longer hold at $O(\alpha_s)$, neither for the
one-loop corrections to the quark handbag diagrams in Fig.~\ref{haba}
nor for the diagrams involving gluon distributions. On general
grounds, the phase structure of the two processes is in fact
different. Whereas the only discontinuity of the two-photon amplitude
in DVCS kinematics is in the $s$-channel, the TCS amplitude has
discontinuities in both $s$ and $Q'^2$, with one-loop hard scattering
diagrams contributing to both cuts. In situations where $O(\alpha_s)$
contributions are important, the DVCS and TCS processes will have a
different dependence on the generalized parton distributions.  TCS and
DVCS together can then constrain them more effectively than either
process alone. The detailed study of TCS at one-loop level is beyond
the scope of this work, and we will base our numerical studies on the
Born level expression (\ref{leading-order}).

\begin{figure}
\begin{center}
  \epsfxsize=0.9\textwidth
  \epsffile{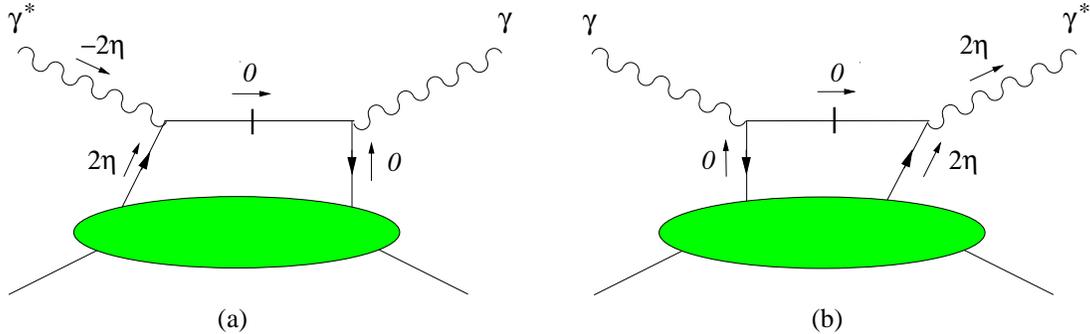}
\caption{\label{fig:poles}The loop momentum configurations $x=\eta$
where the Born level amplitude receives its imaginary part in (a) DVCS
and (b) TCS.  Short vertical lines indicate on-shell quark lines in
the hard scattering, plus-momentum fractions $\pm 2\eta$ and $0$ refer
to the average proton momentum $\frac{1}{2}(p+p')$. The corresponding
configurations for $x=-\eta$ are obtained by reversing the charge flow
of the quark line.}
\end{center}
\end{figure}

It is worthwhile to compare the momentum configurations in DVCS
and TCS from which the Born level convolutions (\ref{htilde}) receive
their imaginary parts. {}From Fig.~\ref{fig:poles} we see that in both
cases there is a quark line with zero plus-momentum coming from the
proton, and that in both cases it is attached to the real photon,
i.e., to the final state in DVCS and to the initial state in TCS.

We conclude this section by defining the variable
\begin{equation}
\tau = \frac{Q'^2}{2 p\cdot q} = \frac{Q'^2}{s - \mm^2}
\end{equation}
for the TCS process as the analog of the Bjorken variable $x_B = Q^2
/(2 p\cdot q)$ in DVCS. The similar roles played by these quantities
reveals itself in their relations with $\eta$, which to leading-twist
accuracy reads $\eta = \tau /(2 - \tau)$ for TCS and $\eta = x_B /(2 -
x_B)$ for DVCS.

\section{The timelike photon}
\label{sec:time}

Processes involving timelike photons can have markedly different
features than processes controlled by large spacelike
virtualities. These features usually do not arise to leading order in
perturbation theory, which is the approximation we will work in
here. A closer look at the Born level diagrams reveals nevertheless
important similarities and differences between timelike processes,
which we now briefly discuss.

\begin{figure}
\begin{center}
   \epsfxsize=0.7\textwidth
   \epsffile{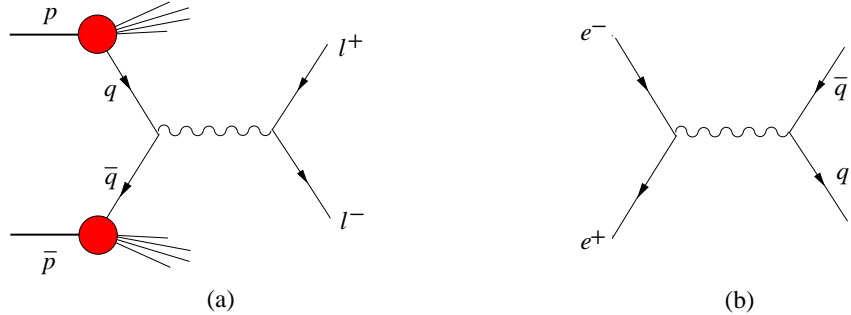}
\caption{\label{fig:timelike} Born level diagrams for the (a)
Drell-Yan process $p\bar{p}\to \ell^+\ell^-\, X$ and (b) $e^+e^-$
annihilation into hadrons, $e^+e^-\to X$.}
\end{center}
\end{figure}

The reaction which at first sight is most similar to TCS is Drell-Yan
pair production in hadron-hadron collisions. In that case, the
$O(\alpha_s)$ corrections to the Born graph of
Fig.~\ref{fig:timelike}a have considerable size and make up for most
of the much discussed $K$-factor of this process. A way to understand
them is the occurrence of large contributions enhanced by $\pi^2$,
which can be traced back to the correction of the quark-photon vertex
for spacelike $\gamma^*$ and on-shell quarks
\cite{Parisi:1980xd}. Notice that in the TCS Born graphs of
Fig.~\ref{haba} only one of the two quark lines attached to the
$\gamma^*$ is on-shell, whereas the other one is off-shell by order
$Q'^2$. One might argue that the second line does become on-shell in
the imaginary part of the amplitude, as indicated in
Fig.~\ref{fig:poles}b, but there is an important difference: the quark
lines in the Drell-Yan diagram and one of the lines in TCS physically
correspond to small virtualities as they are directly attached to
parton distributions, i.e., to quantities describing long-distance
physics. This is not the case for the vertical quark line in the TCS
diagrams. Technically, the singularity of its propagator can be
avoided by analytical continuation of the loop momenta, whereas the
singularities associated with the lines attached to a parton
distribution are pinched \cite{Collins:1999be,Collins:1997fb}. The
analogy between the two processes must hence be used with care, and in
particular one cannot easily infer on the size of the $O(\alpha_s)$
corrections from the experience with the Drell-Yan process.

A second issue in processes with timelike photons is the importance of
resonance effects, which are beyond the realm of perturbation
theory. At invariant photon masses above 4 or 5~GeV, excluding of
course the region of the $\Upsilon$ resonances, the comparison of
leading-twist perturbative calculations and data works rather
satisfactorily for the Drell-Yan process, cf.\ the data compilation in
\cite{Stirling:1993}.  The situation for masses below the $J/\Psi$ is
difficult to assess, mainly due to background lepton pairs from the
weak decays of $b$ and $c$ quarks \cite{Badier:1984ik}. This type of
background does of course not affect TCS, where we are dealing with
\emph{exclusive} lepton pair production. As for inclusive $e^+e^-$
annihilation into hadrons, the recent BES data \cite{Bai:2001ct} in
the mass region from 2 to 3~GeV is remarkably flat and close to the
leading-twist result.  The same holds for the data above 5 GeV, cf.\
e.g.~\cite{Swartz:1996hc}, excluding again the $\Upsilon$ region.
Between 3 and 5 GeV on the other hand, resonance structures are
clearly visible \cite{Bai:2001ct}.

Again one should keep in mind that the importance of resonance effects
may be different in all these processes. In line with our above
analysis, we remark that in the tree level diagram for inclusive
$e^+e^-$ annihilation, Fig.~\ref{fig:timelike}b, both quark and
antiquark correspond to large virtualities. Technically, the cross
section is calculated as the imaginary part of the photon vacuum
polarization, where the quarks appear in a loop and are indeed far
off-shell. We notice that in both Drell-Yan production and $e^+e^-$
annihilation one has quark-antiquark configurations with comparable
virtualities. In contrast, we have asymmetric configurations in TCS,
with one quark line soft and the other far off-shell. Furthermore, the
space-time structure of TCS is such that the $\gamma^*$ is formed from
a $q\bar{q}$-pair only in the ERBL region of the parton distributions,
while in the DGLAP region the parton-level process is photon radiation
off a quark or antiquark, $q\to \gamma^* q$ or $\bar{q}\to \gamma^*
\bar{q}$.

To conclude, we estimate based on $e^+e^-\to X$ and the Drell-Yan data
that ranges of $Q'$ where the leading-twist description of TCS may
work should be between about 1.5 to 2~GeV and the $J/\Psi$ mass, and
above the charmonium resonances. We stress however that the reaction
mechanism in the TCS process displays important differences, and that
one will have to see in the data how parton-hadron duality manifests
itself here.

\section{Observing TCS in lepton pair production}
\label{sec:pair}

\subsection{Some kinematics}
\label{sec:kin}

\begin{figure}[tb]
\begin{center}
  \epsfxsize=0.95\textwidth
  \epsffile{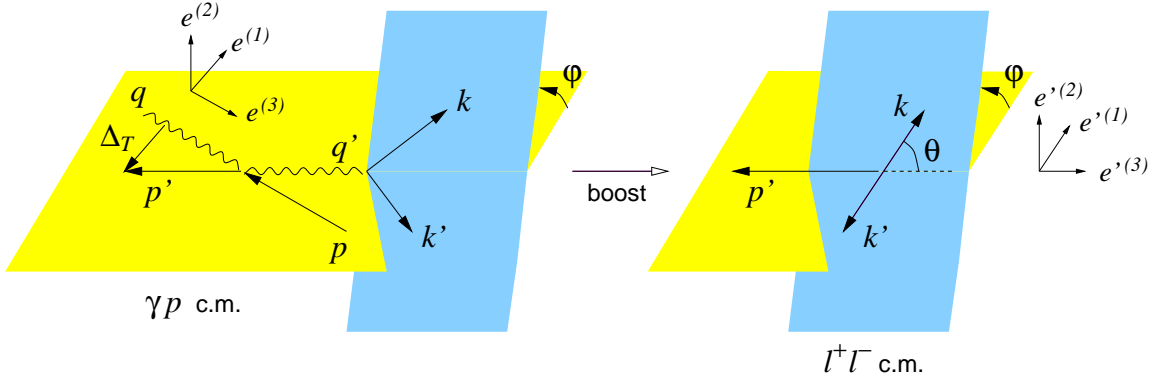}
\caption{Sketch of the kinematical variables and coordinate axes in
the $\gamma p$ and $\ell^+\ell^-$ c.m.\ frames. Notice that the
coordinate systems differ from the one we used in the Compton
amplitude (\protect\ref{leading-order}), where $p$ and $p'$ have
positive $3$-components.}
\label{angle}
\end{center}
\end{figure}

Let us now specify the variables we use to describe the lepton pair
production process depicted in Fig.~\ref{refig}, in addition to those
already introduced at the beginning of
Sect.~\ref{sec:compton}. A~useful quantity is the transverse component
$\vec{\Delta}_T$ of the momentum transfer $\Delta$ with respect to
$\vec{p}$ and $\vec{q}$ in the $\gamma p$ c.m.  It is related to the
scattering angle $\Theta_{\mathrm{cm}}$ in that frame by
\begin{equation}
\sin\Theta_{\mathrm{cm}} = \frac{2 \Delta_T \sqrt{s}}{r} ,
\end{equation}
where $\Delta_T = |\vec{\Delta}_T|$ and $r = \sqrt{ (s-\qq^2-\mm^2)^2
- 4\qq^2\mm^2 }$.  In the limit of large $Q'^2$, large $s$, and small
$-t$, we then have
\begin{equation}
  -t \approx \frac{\tau^2 \mm^2 + \Delta_T^2}{1-\tau}
\label{t-eq}
\end{equation}
up to relative corrections of order $M^2/Q'^2$. For the lepton pair,
we use the lepton velocity
\begin{equation}
 \beta = \sqrt{1 - 4 \ee^2 /\qq^2}
\end{equation}
in the $\ell^+\ell^-$ c.m., where $\ee$ denotes the lepton mass. In
the same frame we introduce the polar and azimuthal angles $\theta$
and $\varphi$ of $\vec{k}$, with reference to a coordinate system with
$3$-axis along $-\vec{p}\,'$ and $1$- and $2$-axes such that $\vec{p}$
lies in the $1$-$3$ plane and has a positive
$1$-component.\footnote{They correspond to the decay angles $\theta$
and $\phi$ for vector meson photoproduction introduced by Schilling,
Seyboth and Wolf \cite{Schilling:1970um} with their vector $\vec{\pi}$
along $\vec{k}$.} This is shown in Fig. \ref{angle}. In terms of
Lorentz invariants, our angles are given by
\begin{eqnarray}
  \label{scalar}
2 (k-k')\cdot p' &=& \beta\, r \cos\theta ,
  \phantom{\frac{1}{r}}
\nonumber \\
2 (k-k')\cdot (p-p') &=& \sigma \beta\,
  \sqrt{(\qq^2-t)^2
         - \left[ \frac{2 (s-\mm^2)\, \qq \Delta_T}{r} \right]^2}\;
  \cos\theta
\nonumber \\
 && {}- \beta\; \frac{2 (s-\mm^2)\, \qq \Delta_T}{r}\,
     \sin\theta \cos\varphi ,
\nonumber \\
4 \epsilon^{\mu\nu\rho\sigma}\,
     p^{\phantom{'}}_\mu p'_\nu k^{\phantom{'}}_\rho k'_\sigma &=&
  \beta\, (s-\mm^2)\, \qq \Delta_T\, \sin\theta \sin\varphi ,
  \phantom{\frac{1}{2}}
\end{eqnarray}
where our convention for the completely antisymmetric tensor is
$\epsilon_{0123} = 1$, and the sign factor $\sigma=\pm 1$ is
determined by
\begin{equation}
\sigma\, \sqrt{(\qq^2-t)^2
         - \left[ \frac{2 (s-\mm^2)\, \qq \Delta_T}{r} \right]^2}\;
  = \frac{Q'^2 (s-\mm^2-Q'^2) + t (s-\mm^2+Q'^2)}{r} .
\end{equation}
The form of the second equation in (\ref{scalar}) is useful in our
kinematics, where $\Delta_T$ is small and $\sigma=1$.

As polarization vectors $\epsilon(\lambda)$ for the incoming photon we
take $\epsilon(\pm) = ( \mp e^{(1)} - i e^{(2)} ) /\sqrt{2}$, where
$e^{(1)}$ and $e^{(2)}$ are unit vectors along the $1$- and
$2$-directions in the $\gamma p$ c.m.\ as shown in
Fig.~\ref{angle}. Our polarizations $\epsilon'(\lambda')$ of the
outgoing photon are $\epsilon'(\pm) = ( \mp e'^{\,(1)} - i e'^{\,(2)}
) /\sqrt{2}$ and $\epsilon'(0) = e'^{\,(3)}$ with unit vectors along
the coordinate axes in the $\ell^+\ell^-$ c.m.\ described above.

%

\subsection{The Bethe-Heitler contribution}
\label{sec:bethe}

\begin{figure}[tb]
\begin{center}
     \epsfxsize=0.8\textwidth
     \epsffile{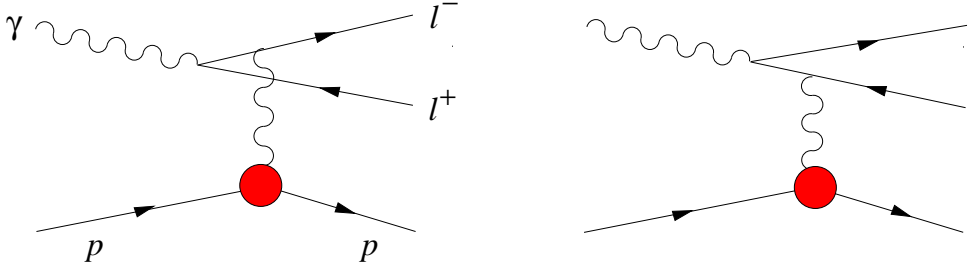}
\caption{The Feynman diagrams for the Bethe-Heitler amplitude.}
\label{bhfig}
\end{center}
\end{figure}

The Bethe-Heitler amplitude is readily calculated from the two Feynman
diagrams in Fig.~\ref{bhfig}. We parameterize the photon-proton vertex
in terms of the usual Dirac and Pauli form factors $F_1(t)$ and
$F_2(t)$, normalizing $F_2(0)$ to be the anomalous magnetic moment of
the target.  We find for the BH contribution to the unpolarized
$\gamma p$ cross section
\begin{equation}
  \label{BH}
\frac{d \sigma_{BH}}{d\qq^2\, dt\, d(\cos\theta)\, d\varphi} =
\frac{\alpha^3_{em}}{4\pi (s-\mm^2)^2}\; \frac{\beta}{-t L}
  \left[ \Big(F_1^2 -\frac{t}{4\mm^2} F_2^2\Big) \frac{A}{-t}
         + (F_1+F_2)^2\, \frac{B}{2} \,\right] ,
\end{equation}
where we have used the abbreviations
\begin{eqnarray}
A &=& (s-\mm^2)^2 \Delta_T^2 - t\, a(a+b) - \mm^2 b^2 
      - t\, (4\mm^2 -t) Q'^2
\nonumber \\
 && {}+ \frac{\ee^2}{L} \left[
           \Big\{ (Q'^2-t)(a+b) - (s-\mm^2)\,b \,\Big\}^2 +
           t\, (4\mm^2 - t) (Q'^2-t)^2 \right]
  \nonumber \\
B &=& (Q'^2+t)^2 + b^2 + 8\ee^2 Q'^2
      - \frac{4\ee^2 (t+2\ee^2)}{L}\, (Q'^2-t)^2 .
\label{BH-coeff}
\end{eqnarray}
The cross section depends on the angles $\theta$ and $\varphi$ through
the scalar products
\begin{equation}
a = 2 (k-k')\cdot p' , \hspace{3em}
b = 2 (k-k')\cdot (p-p')
\end{equation}
given in Eq.~(\ref{scalar}) above, and through the product of the
lepton propagators in the two BH diagrams,
\begin{equation}
  \label{prop}
  L = \Big[ (q-k)^2 - \ee^2 \Big]\, \Big[ (q-k')^2 - \ee^2 \Big]
    = \frac{(\qq^2-t)^2 - b^2}{4} .
\end{equation}

These expressions are rather lengthy, but simplify considerably in
kinematics where $t$, $\mm^2$ and $\ee^2$ can be neglected compared to
terms going with $s$ or $\qq^2$.  We then have
\begin{equation}
  \label{approx-prop}
  L \approx L_0 = \frac{\qq^4 \sin^2\theta}{4} .
\end{equation}
and
\begin{equation}
  \label{approx-BH}
\frac{d \sigma_{BH}}{d\qq^2\, dt\, d(\cos\theta)\, d\varphi}
  \approx
    \frac{\alpha^3_{em}}{2\pi s^2}\, \frac{1}{-t}\,
    \frac{1 + \cos^2\theta}{\sin^2\theta} \,
 \left[ \Big(F_1^2 -\frac{t}{4\mm^2} F_2^2\Big)
            \frac{2}{\tau^2}\, \frac{\Delta_T^2}{-t}\,
        + (F_1+F_2)^2 \,\right] .
\end{equation}
We see that the product $L$ of lepton propagators goes to zero at
$\sin\theta=0$ in this approximation. Closer inspection reveals that
when $\sin\theta$ becomes of order $\Delta_T /\qq$ or $\ee /\qq$ the
approximations (\ref{approx-prop}) and (\ref{approx-BH}) break down
and one must use the full expressions.

Let us see how small the product $L$ can become. At fixed $s$,
$\qq^2$, $t$, $\varphi$ we find with Eqs.~(\ref{scalar}) and
(\ref{prop}) that $L$ assumes a minimum value
\begin{equation}
L_{\mathrm{min}} \approx
  \qq^2 \ee^2 + \qq^2 \Delta_T^2\, \frac{\sin^2\varphi}{(1-\tau)^2}
\end{equation}
for
\begin{equation}
  \tan\theta_{\mathrm{min}}
  \approx - \frac{2 \Delta_T}{\qq} \, \frac{\cos\varphi}{1-\tau} ,
\end{equation}
up to corrections of order $t /Q'^2$, $\mm^2 /Q'^2$, $\ee^2 /Q'^2$.
For $\theta \sim \theta_{\mathrm{min}}$ the leptons $\ell^-$ and
$\ell^+$ are nearly collinear with the initial photon in the $\gamma
p$ c.m.  They have transverse momenta of order $\Delta_T$ with respect
to $\vec{p}$ and $\vec{q}$ and share their total longitudinal momentum
in a highly asymmetric way. In our numerical studies we will impose a
cut on $\theta$ which ensures that $L$ remains of order $Q'^4$, thus
staying away from the region where the BH cross section becomes
extremely large.

We finally remark that as long as $L$ is of order $Q'^4$ the terms
going with $1/L$ in (\ref{BH-coeff}) are suppressed at least like
$\ee^2\, Q'^2/L$ compared with the leading behavior of $A$ and $B$.
For a large range in $\theta$ the BH cross section (\ref{BH}) will
thus approximately behave like $1/L$ instead of $1/L^2$.

\subsection{The Compton scattering contribution}
\label{sec:coco}

We now investigate the TCS contribution to lepton pair production. In
order to understand the basics of its interplay with the BH process it
is sufficient to consider the leading behavior of the Compton
amplitude in $1/Q'$ and in $\alpha_s$, which we discussed in
Sect.~\ref{sec:compton}. We will thus in particular discard $\gamma p$
amplitudes that change the photon helicity. In line with neglecting
power suppressed effects in the Compton subprocess we will also drop
mass corrections of order $M^2/Q'^2$ and $m_\ell^2/Q'^2$ in kinematics
and phase space. To this accuracy, the contribution of TCS to the
unpolarized cross section of $\gamma p\to \ell^+\ell^-\, p$ reads
\begin{equation}
  \label{TCS}
  \frac{d \sigma_{TCS}}{d\qq^2\, dt\, d(\cos\theta)\, d\varphi} \approx
  \frac{\alpha^3_{em}}{8\pi s^2}\,
  \frac{1}{\qq^2}\, \frac{1 + \cos^2\theta}{4}
  \sum_{\lambda,\lambda'} | M^{\lambda'-,\lambda-} |^2 .
\end{equation}
We note that the $\varphi$ independence here is a consequence of
having neglected photon helicity changing transitions. From
(\ref{leading-order}) we obtain
\begin{eqnarray}
\label{squared-amps}
\frac{1}{2}
\sum_{\lambda,\lambda'} | M^{\lambda'-,\lambda-} |^2
 &=& (1-\eta^2) \Big( |{\cal H}_1|^2 + |\tilde{\cal H}_1|^2 \Big)
      - 2 \eta^2\, \mbox{Re} 
      \Big( {\cal H}_1^*\, {\cal E}_1^{\phantom{*}}
          + \tilde{\cal H}_1^*\, \tilde{\cal E}_1^{\phantom{*}} \Big)
\nonumber \\
 && {} - \Big( \eta^2 + \frac{t}{4 M^2} \Big)\, |{\cal E}_1|^2
       - \eta^2 \frac{t}{4 M^2}\, |\tilde{\cal E}_1|^2 ,
\end{eqnarray}
where ${\cal H}_1$, $\tilde{\cal H}_1$, ${\cal E}_1$, $\tilde{\cal
E}_1$ are to be evaluated at $-\xi=\eta$. Together with
Eq.~(\ref{approx-BH}) we see that compared with the TCS cross
section, the BH contribution is parametrically enhanced by a factor
$Q'^2 /(-t)$ and has an extra factor of $1 /\sin^2\theta$ in the
angular dependence.

Let us compare the TCS result (\ref{TCS}) with the contribution of
DVCS to the electroproduction process
\begin{equation}
\ell(k) + p(p) \to \ell(k') + \gamma(q') + p(p') ,
\end{equation}
where we have indicated four-momenta in parentheses. Retaining only
the leading part in $1/Q$ and $\alpha_s$ of the Compton amplitude, and
dropping again mass corrections of order $M^2/Q^2$ and $m_\ell^2/Q^2$,
we have for the unpolarized cross section
\begin{equation}
  \label{DVCS}
  \frac{d \sigma_{DVCS}}{dQ^2\, dt\, dy\, d\varphi} \approx
  \frac{\alpha^3_{em}}{8\pi s_{ep}^2}\,
   \frac{1}{Q^2}\, \frac{1+(1-y)^2}{y^3}
  \sum_{\lambda,\lambda'} | M^{\lambda'+,\lambda+} |^2 .
\end{equation}
Here $s_{ep}=(p+k)^2$ is the total c.m.\ energy of the $ep$ collision,
$y = (q\cdot p) / (k\cdot p)$ the usual inelasticity parameter, and
$\varphi$ the azimuthal angle between lepton and hadron planes as
defined in \cite{Diehl:1997bu}. With the relation (\ref{TCS-DVCS-amp})
we readily see that to leading twist and leading order in $\alpha_s$
the sums over squared helicity amplitudes in (\ref{TCS}) and
(\ref{DVCS}) give identical results for corresponding values of
$\eta=\tau/(2-\tau)$ and $Q'^2$ in TCS, and $\eta=x_B/(2-x_B)$ and
$Q^2$ in DVCS. To this accuracy, the Compton scattering contributions
to the respective cross sections only differ by the global kinematic
factors given in (\ref{TCS}) and (\ref{DVCS}).

Comparison of these factors reveals the correspondence between the
variables $\theta$ in TCS and $y$ in DVCS, which by expressing them in
terms of scalar products is found to be
\begin{eqnarray}
  \label{correspond}
\frac{1+\cos\theta}{2} \approx \frac{k\cdot p'}{(k+k') \cdot p'}
 & \leftrightarrow &
\frac{k\cdot p}{(k-k') \cdot p} = \frac{1}{y} ,
\end{eqnarray}
where in the first relation we have again neglected mass
corrections. At this point we find a crucial difference in the
phenomenology of the two processes. As is well known
\cite{Diehl:1997bu,Kroll:1996pv} the relative weight of DVCS and BH
crucially depends on $y$, given that at amplitude level the DVCS
contribution comes with a factor $1/y$ relative to the BH
contribution. In the region of $Q^2$ and $t$ defining the DVCS regime,
BH dominates for moderate values of $y$, whereas for sufficiently
small values of $y$ the Compton contribution wins. Since the quantity
corresponding to $1/y$ in Eq.~(\ref{correspond}) is always between
$-1$ and 1, no such enhancement takes place for TCS, and we will
indeed find numerically that here the BH contribution to the cross
section is always dominant. The strategy is then the same as in DVCS
at moderate values of $y$, namely to gain information on the Compton
process through its interference with BH, which can be extracted using
symmetry properties of the process.

Another noteworthy difference concerns the variables $\tau$ and $x_B$,
which determine the values $\eta$ where the generalized parton
distributions are probed in the two processes. In DVCS at fixed
collision energy $\sqrt{\rule{0pt}{1.5ex} s_{ep}}$ the variables $x_B$
and $y$ are not independent since $Q^2 = y x_B\,
(s_{ep}-\ee^2-\mm^2)$. If at given $Q^2$ one wants to vary $x_B$ and
thus choose a kinematical point where to probe the Compton subprocess,
one must vary $y$. In TCS on the other hand, one has the relation
$Q'^2 = \tau\, (s- \mm^2)$, independent of the value of~$\theta$. In
order to vary $\tau$ at given $Q'^2$, one here needs to change the
$\gamma p$ collision energy $\sqrt{s}$. A continuous spectrum in $s$
is of course automatically obtained if the initial photon originates
from bremsstrahlung off a lepton beam.

\subsection{The interference term}
\label{sec:inter}

Let us now explore how information on the Compton process can be
obtained from the interference between the TCS and BH amplitudes. The
general strategy is the same as described in \cite{Diehl:1997bu} for
the case of DVCS, but we will again encounter important differences in
the phenomenology of these reactions.

A key point is that the amplitudes for the Compton and Bethe-Heitler
processes transform with opposite signs under reversal of the lepton
charge. As a consequence the interference term between TCS and BH is
odd under exchange of the $\ell^+$ and $\ell^-$ momenta, whereas the
individual contributions of the two processes are even. Any observable
that changes sign under $k\leftrightarrow k'$ will hence project out
the interference term, eliminating in particular the large BH
contribution. Clean information on the interference term is therefore
contained in the angular distribution of the lepton pair. The
corresponding observable in the electroproduction process $\ell p\to
\ell \gamma p$ is the lepton beam charge asymmetry, whose measurement
presents important experimental challenges.

Let us take a closer look at the interference part of the cross
section for $\gamma p\to \ell^+\ell^-\, p$ with unpolarized protons
and photons. It is given by
\begin{eqnarray}
  \label{intres}
\frac{d \sigma_{INT}}{d\qq^2\, dt\, d(\cos\theta)\, d\varphi}
&=& {}-
\frac{\alpha^3_{em}}{4\pi s^2}\, \frac{1}{-t}\, \frac{\mm}{Q'}\,
\frac{1}{\tau \sqrt{1-\tau}}\, \frac{L_0}{L}
\left[\, \cos\varphi\, \frac{1+\cos^2\theta}{\sin\theta}\,
    \re\tilde{M}^{--}
\right.
\nonumber \\
&& \left. \hspace{-4em} {}
  - \cos2\varphi\, \sqrt{2}\cos\theta\, \re\tilde{M}^{0-}
  + \cos3\varphi\, \sin\theta\, \re\tilde{M}^{+-}
+ O\Big( \frac{1}{Q'} \Big)
\right] ,
\end{eqnarray}
with $L$ and $L_0$ from Eqs.~(\ref{prop}) and (\ref{approx-prop}).
Here
\begin{eqnarray}
\tilde{M}^{\mu'\mu} &=& \frac{\Delta_T}{M}
    \left[ (1-\tau) F_1 - \frac{\tau}{2}\, F_2 \,\right]
       M^{-\mu',-\mu}
+ \frac{\Delta_T}{M} \left[ F_1 + \frac{\tau}{2}\, F_2 \,\right]
       M^{+\mu',+\mu}
\nonumber \\
&& {}+ \left[ \tau^2 (F_1+F_2) + \frac{\Delta_T^2}{2M^2} F_2 \,\right]
       M^{-\mu',+\mu}
- \frac{\Delta_T^2}{2M^2}\, F_2\, M^{+\mu',-\mu}
\end{eqnarray}
is the same combination of Compton helicity amplitudes as defined in
\cite{Diehl:1997bu}.\footnote{In contrast to
\protect\cite{Diehl:1997bu} our notation here is to list the
helicities of outgoing particles first. With our phase convention the
transverse polarization vectors of the two photons coincide for
$\Delta_T=0$, cf.\ Sect.~\ref{sec:kin}. In~\protect\cite{Diehl:1997bu}
we made a different choice, and the Compton helicity amplitudes here
and there differ by an overall sign.}
The close analogy between TCS and DVCS is manifest, and we see that a
$\gamma^*$ with negative helicity in TCS corresponds to a $\gamma^*$
with positive helicity in DVCS as we already found in the relations
(\ref{TCS-DVCS-amp}).

The terms indicated by $O(1/Q')$ in Eq.~(\ref{intres}) have
kinematical coefficients suppressed by at least one power of $1/Q'$
relative to the other terms in brackets. Notice that we have not
approximated the product $L$ of lepton propagators from the BH
process. In the limit of large $Q'^2$ the factor $L_0 /L$ tends to 1,
but we have seen in Sect.~\ref{sec:bethe} that this approximation
becomes increasingly bad as $\theta$ approaches $0$ or $\pi$, so that
it is useful to keep $L_0 /L$ in an analysis. The same is true for the
lepton propagators in the interference of DVCS and BH, as has been
emphasized in \cite{Belitsky:2001gz}.

We see that without polarization one probes the real parts of the
Compton helicity amplitudes. Access to the imaginary parts can be
obtained with polarized photon beams. If the photons have a circular
polarization $\nu$, as is the case for a bremsstrahlung beam emitted
from longitudinally polarized leptons, one has
\begin{eqnarray}
  \label{helasy} 
\frac{d \sigma_{INT}}{d\qq^2\, dt\, d(\cos\theta)\,d\varphi} 
&=& \left. 
\frac{d \sigma_{INT}}{d\qq^2\, dt\, d(\cos\theta)\, d\varphi}
\right|_{\rm eq.\,(\protect\ref{intres})} 
\nonumber \\ &&
\hspace{-4em} {}-
\nu\; \frac{\alpha^3_{em}}{4\pi s^2}\, \frac{1}{-t}\, 
\frac{\mm}{Q'}\, \frac{1}{\tau \sqrt{1-\tau}}\, \frac{L_0}{L} 
\left[\, \sin\varphi\,
\frac{1+\cos^2\theta}{\sin\theta}\, \im\tilde{M}^{--} \right.
\nonumber \\ 
&& \left. \hspace{-4em} {} - \sin2\varphi\,
\sqrt{2}\cos\theta\, \im\tilde{M}^{0-} + 
\sin3\varphi\, \sin\theta\, \im\tilde{M}^{+-} 
+ O\Big( \frac{1}{Q'} \Big) \right] .
\end{eqnarray}
The photon polarization dependent and independent terms are simply
related by exchanging $\sin \leftrightarrow \cos$ and $\im
\leftrightarrow \re$. This is not quite the same as for \emph{lepton}
beam polarization in the interference between DVCS and BH, where
different kinematical factors occur in the polarization dependent and
independent parts, and where notably the term with $\sin3\varphi\,
\im\tilde{M}^{+-}$ is absent.

The various terms in the $\varphi$ dependence of the interference term
can for instance be projected out by weighting the differential cross
section with appropriate functions. The weights $(L /L_0)
\cos(n\varphi)$ and $(L /L_0) \sin(n\varphi)$ for instance give the
terms with $\cos(n\varphi)$ and $\sin(n\varphi)$ in Eq.~(\ref{intres})
and (\ref{helasy}), respectively. Notice that these weights are odd
under the exchange of $k$ and $k'$ and hence do not pick up the BH and
TCS contributions to the cross section, as discussed above.

In this way we can project out the various helicity combinations
$\tilde{M}^{\mu'\mu}$ of Compton amplitudes, up to relative
corrections in $1/Q'$.  Along the lines of \cite{Diehl:1997bu} this
can be used to test whether the power behavior in $Q'$ at fixed $\tau$
and $t$ follows the predictions discussed in Sect.~\ref{sec:compton},
i.e., whether arguments based on the large $Q'^2$ limit apply at the
finite $Q'^2$ of a measurement.  If one is in the scaling regime, one
can then analyze the photon helicity conserving amplitudes in terms of
generalized parton distributions. The quark handbag diagrams of
Fig.~\ref{haba} give
\begin{equation}
\label{mmimi}
\tilde{M}^{--} = \frac{2\sqrt{t_0-t}}{M}\, \frac{1-\eta}{1+\eta}\,
\left[ F_1 {\cal H}_1 - \eta (F_1+F_2)\, \tilde{\cal H}_1 -
\frac{t}{4M^2} \, F_2\, {\cal E}_1 \,\right]
\end{equation}
where $-t_0 = 4\eta^2 M^2 /(1-\eta^2)$ is the minimal value of $-t$ at
given $\eta$, up to corrections in~$1/Q'^2$.

The above extraction of the the Compton amplitudes requires
measurement of the angle $\varphi$. If one integrates the interference
term over $\varphi$, the photon polarization dependent part in
(\ref{helasy}) vanishes because of parity invariance.  The integral of
the unpolarized contribution (\ref{intres}) is nonzero, due to the
$\varphi$ dependence of $L_0/L$ and to the $\varphi$ independent part
of the terms denoted by $O(1/Q')$.  This integral can in principle be
projected out from the cross section because it is odd under $\theta
\to \pi-\theta$, whereas the BH and TCS contributions are even when
integrated over $\varphi$.  The interference signal so obtained is
however an order $1/Q'$ smaller than what can be seen in the $\varphi$
dependence of the cross section, and will thus be harder to extract.

\section{Numerical estimates}
\label{sec:numbers}

In this section we model the generalized parton distributions (GPDs)
and give estimates for various observables. We restrict ourselves to
moderate values of $\tau$ and use the leading-order handbag
approximation (\ref{leading-order}), (\ref{htilde}) of the Compton
amplitude.  We omit all terms proportional to $E^q$ and
$\tilde{E}^q$. In the region $0.1\le\tau\le 0.36$ and $|t|\le
0.4$~GeV$^2$ we will consider in our estimates, ${\cal E}_1$ is
multiplied by kinematical coefficients at most $0.15$ times those of
${\cal H}_1$ in (\ref{squared-amps}) and (\ref{mmimi}) and thus would
not significantly change our results. In any case, it is at present
fairly unclear how to model the distributions $E^q$, so that taking
them into account would not improve the reliability of the
estimates. As for $\tilde{\cal E}_1$, it is multiplied by a tiny
coefficient in (\ref{squared-amps}) and absent in the interference
term (\ref{mmimi}).

\subsection{Modeling the parton distributions}
\label{sec:model}

Now we define the model we use for $H^q$ and $\tilde{H}^q$. Following
\cite{Guichon:1998xv} we take a factorizing ansatz for the $t$
dependence,
\begin{eqnarray}
H_{D\!D}^u(x,\eta,t) &=& h^u(x,\eta)\, {\textstyle\frac{1}{2}}
                         F_{1}^u(t),
  \nonumber \\
H_{D\!D}^d(x,\eta,t) &=& h^d(x,\eta)\,F_{1}^d(t),
  \nonumber \\
\tilde{H}_{\dd}^q(x,\eta,t) &=& \tilde{h}^q(x,\eta)\, \tilde{F}^q(t),
\label{gpd-factor}
\end{eqnarray}
with
\begin{eqnarray}
  \label{form}
F_{1}^u(t) &=& 2 F_1^p(t) + F_1^n (t),
\nonumber \\
F_{1}^d(t) &=& F_1^p(t) + 2 F_1^n(t) ,
\nonumber \\
\tilde{F}^u(t) \;=\; \tilde{F}^d(t) &=&  g_A(t)/g_A(0) .
\end{eqnarray}
$F_1^p$ and $F_1^n$ are the electromagnetic Dirac form factors of the
proton and neutron, for which we take the usual dipole
parameterization \cite{Boffi:1993gs}. For the axial form factor of the
proton we take $g_A(t) = g_A(0)\, (1 - t /M^2_A)^{-2}$ with
$g_A(0)=1.26$ and $M_A = 1.06$~GeV from~\cite{Ahrens:1987xe}.  For
strange quarks we make the ansatz
\begin{eqnarray}
H_{D\!D}^s(x,\eta,t) &=& h^s(x,\eta)\, F_D(t),
\label{gpd-strange}
\end{eqnarray}
where the dipole form factor $F_D(t) = (1 - t /M^2_V)^{-2}$ with $M_V
= 0.84$~GeV is the same that enters in the parameterization of $F_1^p$
and $F_1^n$. Note that via the sum rule for $\int dx\, H^s(x,\eta,t)$
a factorizing ansatz like (\ref{gpd-strange}) corresponds to setting
the strange quark contribution $F_{1}^s(t)$ to the Dirac form factor
to zero.  We remark that several studies
\cite{Penttinen:2000th,Diehl:1999kh} indicate that GPDs do \emph{not}
factorize in the simple manner of (\ref{gpd-factor}) and
(\ref{gpd-strange}). The ansatz has however the virtue of simplicity
and should be good enough for our estimates, as long as we do not
study the interplay of the $\eta$ and $t$ dependence of the cross
section.  For $h^q$ and $\tilde{h}^q$ we make an ansatz based on
double distributions~\cite{Musatov:2000xp},

\begin{eqnarray}
h^q(x,\eta) &=& \int_{0}^{1} dx'
  \int_{-1+x'}^{1-x'} dy' \,
  \Bigg[
  \delta(x-x'- \eta y') \, q(x')
\nonumber\\
&& \hspace{6.8em} {}-
  \delta(x+x' - \eta y') \, \bar{q}(x') \,
  \Bigg] \, \pi(x',y') ,
\label{ddmodel-h}
\\
\tilde{h}^q(x,\eta)&=&\int_{0}^{1} dx'
  \int_{-1+x'}^{1-x'} dy' \,
  \delta(x-x' - \eta y')\, \Delta q_V(x')\, \pi(x',y') ,
\label{ddmodel-ht}
\\
\pi(x',y') &=& \frac{3}{4}\,
  \frac{(1-x')^2 - y'^2}{(1-x')^3} .
\label{profile}
\end{eqnarray}
We evaluate (\ref{ddmodel-h}) with the LO GRV 94 parameterization
\cite{Gluck:1995uf} of the unpolarized distributions $q(x)$ and
$\bar{q}(x)$, and (\ref{ddmodel-ht}) with set A of the LO polarized
valence distributions $\Delta q_V(x)$ by Gehrmann and
Stirling~\cite{Gehrmann:1996ag}. In both cases we take the
factorization scale as $\mu^2 = 5 \gev^2$. We neglect the polarized
quark sea, which presently is not well constrained by data, and
thereby also drop $\tilde{H}^s$.  In the appendix we shall give a
detailed discussion of the role played by very small values of $x'$ in
the integrals of (\ref{ddmodel-h}) and (\ref{ddmodel-ht}), and thus of
the uncertainties in evaluating them with parton densities only known
above some finite value of $x'$.

Let us stress that the available models of GPDs are fraught with
uncertainties, in particular in the ERBL region. There, GPDs describe
the emission of a $q\bar{q}$ pair from the target, and an ansatz only
using the information from usual parton densities should be used with
care.  Dynamical calculations \cite{Petrov:1998kf,Penttinen:2000th}
lead in fact to much richer structure in the ERBL region than is
generated from (\ref{ddmodel-h}) to (\ref{profile}).  Notice also
that, while for $x>\eta$ GPDs are bounded from above
\cite{Pire:1999nw}, no analogous constraints are known in the ERBL
region.

A particular type of contribution in the ERBL region is the
Polyakov-Weiss $D$-term \cite{Polyakov:1999gs}, which following
\cite{Kivel:2001fg} we take as a flavor $SU(3)$ singlet
\begin{eqnarray}
  \label{dt}
H^u_{D}(x,\eta,t) = H^d_{D}(x,\eta,t) = H^s_{D}(x,\eta,t) =
   \Theta(\eta^2 - x^2)\, \frac{1}{3}
   D\Big(\frac{x}{\eta}\Big)\, F_D(t),
\end{eqnarray}
where $\Theta$ denotes the step function.  We make again a factorizing
ansatz for the $t$ dependence, taking the same dipole form factor as
in (\ref{gpd-strange}).  For the function $D$ we use the
parameterization given in equations~(23) and (24) of
\cite{Kivel:2001fg}, which was obtained by a fit to the result
obtained in the chiral soliton model~\cite{Petrov:1998kf}.  That
parameterization is given for a factorization scale $\mu= 0.6 \gev$,
and we use the leading-order evolution equations to evolve it
up. Because of mixing we then need the $D$-term in the gluon GPD of
the proton, which we take as zero at $\mu= 0.6 \gev$.  Following
\cite{Gluck:1995uf} we take $\Lambda^{(3)} = 232 \mbox{~MeV}$ and
$\Lambda^{(4)} = 200 \mbox{~MeV}$ for the scale parameter in
$\alpha_S$, switching from 3 to 4 flavors at $\mu=1.5 \gev$.  For
$\mu^2 = 5 \gev^2$ we then get
\begin{equation}
\label{dt-at-5}
D(z) \approx {}- (1-z^2) \Big[\, 2.9\, C_1^{3/2}(z) 
            + 0.6\, C_3^{3/2}(z) + 0.2\, C_5^{3/2}(z) \,\Big]
\end{equation}
with Gegenbauer polynomials $C_n^{3/2}(z)$.  Below, we will give
estimates with and without the $D$-term contribution according to
(\ref{dt}) and (\ref{dt-at-5}) in order to explore the model
dependence of our results.

%
\begin{figure}
\begin{center}
\leavevmode
   \epsfxsize=0.49\textwidth
     \epsffile{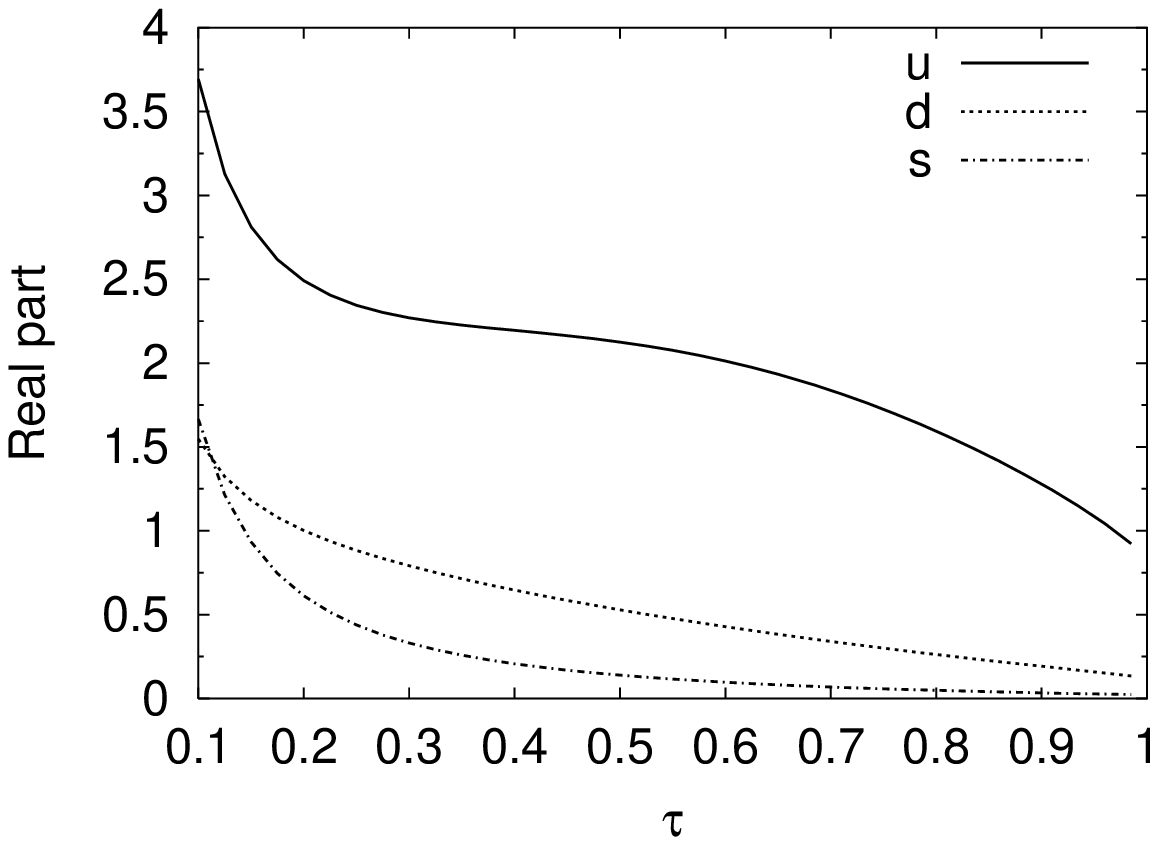}
   \epsfxsize=0.49\textwidth
     \epsffile{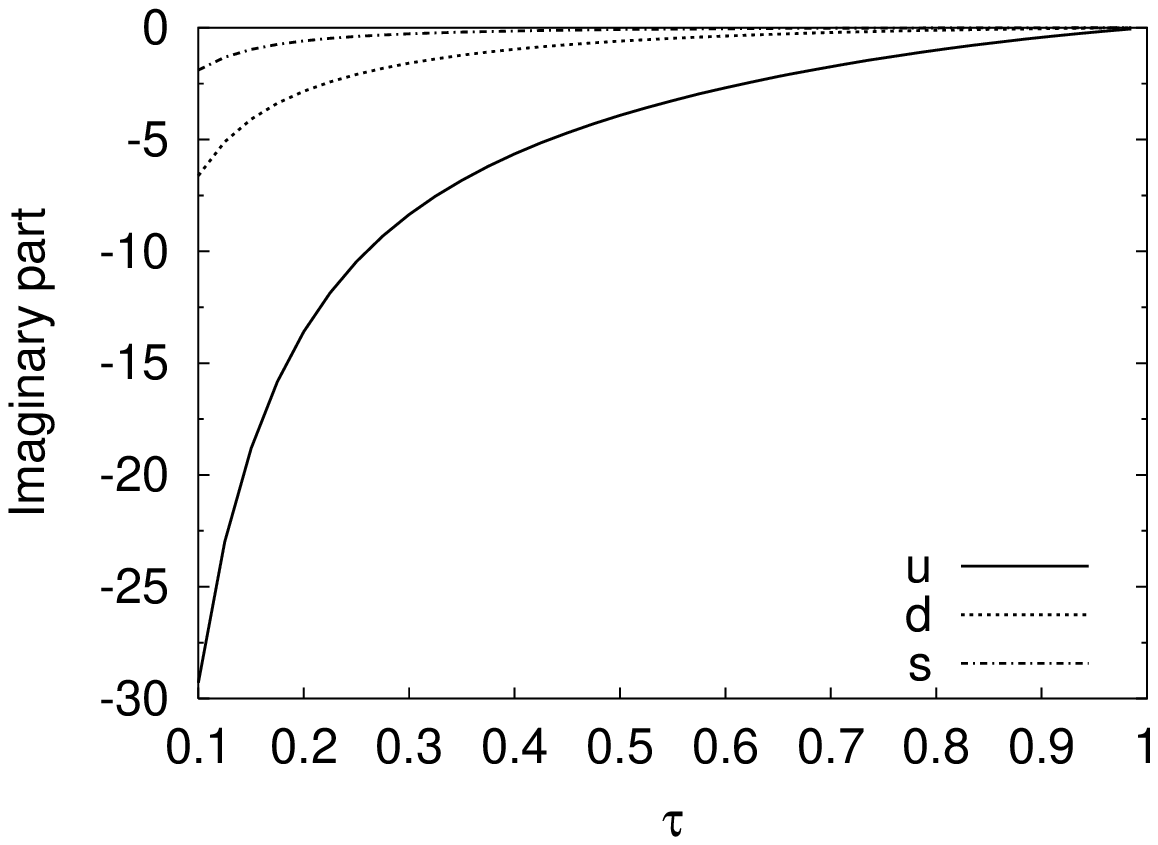}
\caption{The contributions from $u$, $d$, and $s$ quarks to $\re{\cal
H}_1$ (left) and $\im{\cal H}_1$ (right). They are calculated with
$H^q_{\protect\phantom{!}} = H_{\dd}^q$ and respectively divided by
$\frac{1}{2} F_1^u(t)$, $F_1^d(t)$, and $F_D(t)$.}
\label{hfig}
\end{center}
\begin{center}
   \epsfxsize=0.49\textwidth
     \epsffile{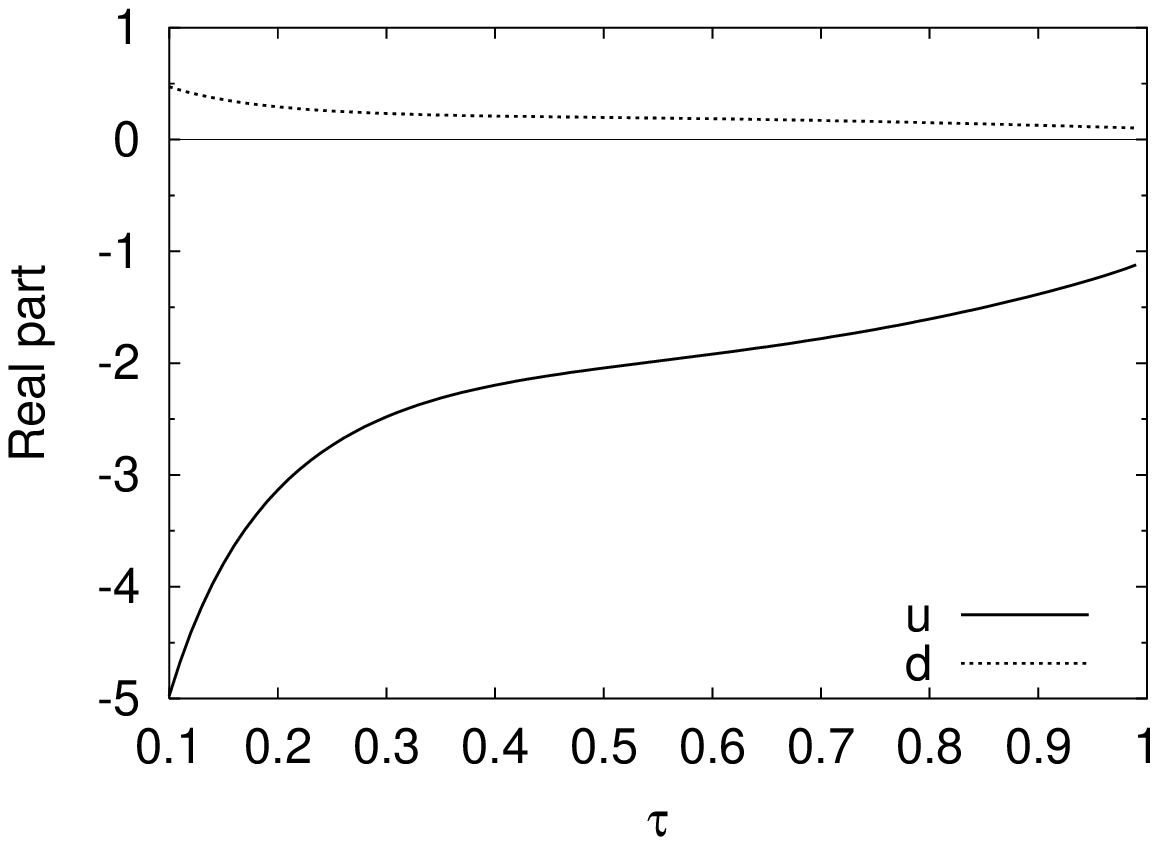}
   \epsfxsize=0.49\textwidth
     \epsffile{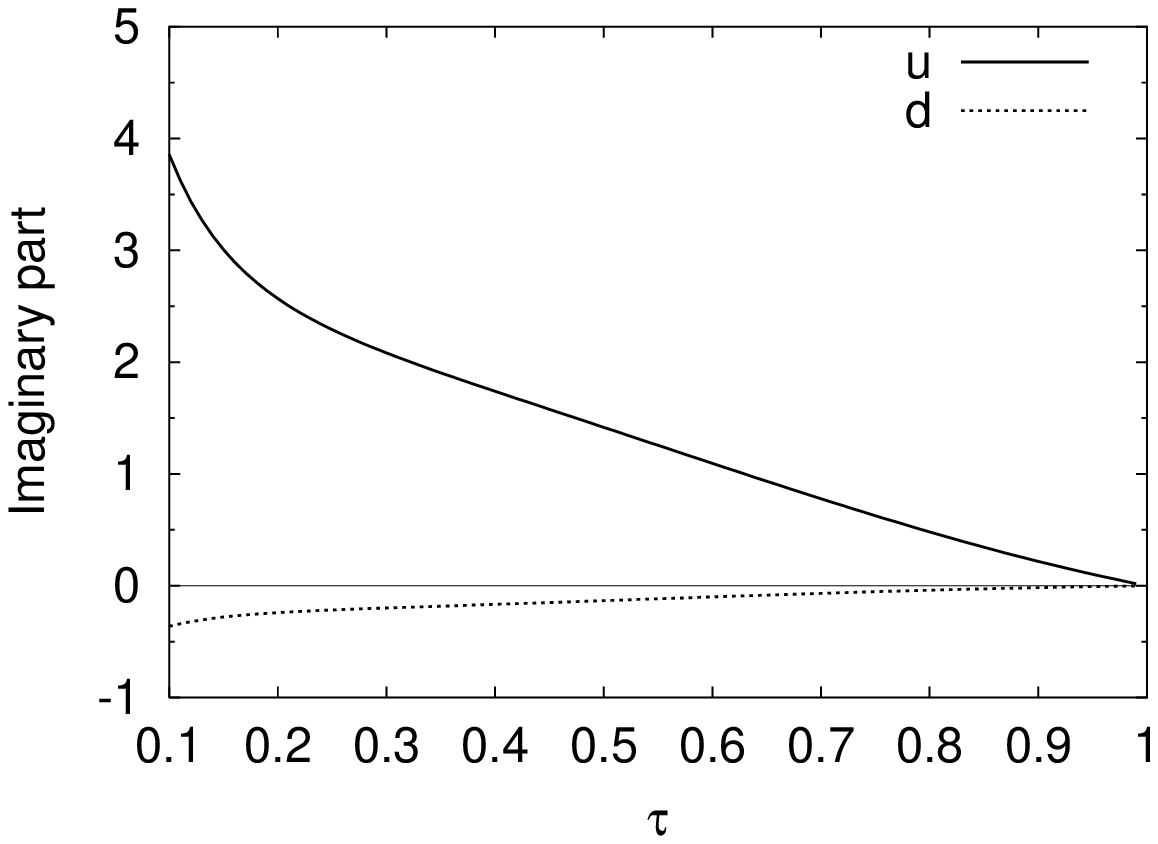}
\caption{The contributions from $u$ and $d$ quarks to $\re\tilde{\cal
H}_1$ (left) and $\im\tilde{\cal H}_1$ (right). They are calculated
with our model for $\tilde{H}^q$ and divided by $\tilde{F}^q(t)$.}
\label{htfig}
\end{center}
\end{figure}
%

In Fig.~\ref{hfig} we show the real and imaginary parts of the
convolution integral ${\cal H}_1(-\eta,\eta,t)$, calculated from
$H_{\dd}$.  Decomposing ${\cal H}_1 = {\cal H}_1^{u} + {\cal H}_1^{d}
+ {\cal H}_1^{s}$ we plot the contributions from $u$, $d$, and $s$
quarks separately. We further divide by appropriate factors
$\frac{1}{2} F_1^u(t)$, $F_1^d(t)$, and~$F_D(t)$, so that with the
factorizing ansatz (\ref{gpd-factor}), (\ref{gpd-strange}) the
resulting curves are independent of $t$.\footnote{Formally, these
curves correspond to ${\cal H}_1^{u,d,s}$ at $t=0$, which for
$\tau\neq 0$ is however outside the physical region according to
(\protect\ref{t-eq}).}
Analogous plots for $\tilde{\cal H}_1 = \tilde{\cal H}_1^{u} +
\tilde{\cal H}_1^{d}$ are given in Fig.~\ref{htfig}.  We observe that
for $\tau \sim 0.1$ the $s$ quark contribution to $\re{\cal H}_1$ is
by no means small compared with $u$ and $d$ quarks, although it is
tiny in $\im{\cal H}_1$.  This illustrates that at least the real part
of the Compton amplitude is not related in a straightforward manner
with the usual parton densities at $x\sim \tau$, given that $s(x)$
only becomes comparable to $u(x)$ and $d(x)$ for $x$ considerably
below 0.1.  

\begin{figure}
\begin{center}
   \epsfxsize=0.49\textwidth
     \epsffile{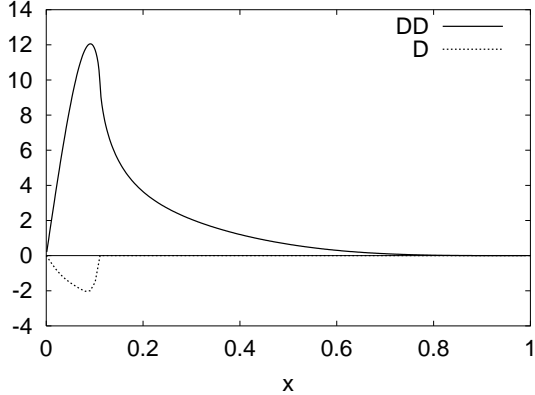}
\caption{The combination $H_{\dd}^u(x,\eta,t) - H_{\dd}^u(-x,\eta,t)$
divided by $\frac{1}{2} F_1^u(t)$, and the combination
$H_D^u(x,\eta,t) - H_D^u(-x,\eta,t)$ divided by $F_D(t)$.  Both curves
are for $\eta=0.11$, corresponding to $\tau=0.2$.}
\label{ddfig}
\end{center}
\end{figure}

We do not show in Fig.~\ref{hfig} the contributions from $H_D$. They
are only nonzero in the real part, summed over all flavors they amount
to a $\tau$ independent contribution of $-3.3 \, F_D(t)$ in ${\cal
H}_1$.  The remarkable fact that the $D$-term contribution to the TCS
amplitude is independent of $\eta$ at fixed $t$ remains true to all
orders in perturbation theory. This is because due to general scaling
properties the hard scattering kernel can be written as the
leading-order one in (\ref{htilde}) times a function of $x /\eta$.
Comparing with Fig.~\ref{hfig} we see that the $D$-term has an
appreciable impact on the value of $\re{\cal H}_1$ in our model.  This
is surprising if one compares the functions $H_{D}$ and $H_{\dd}$
themselves.  We show this for $u$ quarks in Fig.~\ref{ddfig}, plotting
only the charge conjugation even combination $H(x,\eta,t) -
H(-x,\eta,t)$ that enters in Compton scattering.  One can understand
the strong amplification of a moderate change in the ERBL region of a
GPD by observing that the real part in the convolutions (\ref{htilde})
is a principal value integral, which involves large cancellations
between the contributions from $|x|<\eta$ and $|x|>\eta$.  Here is one
of the reasons why measuring the real part of the Compton amplitude,
in DVCS or in TCS, can provide unique information on generalized
parton distributions.

%
%

\subsection{Cross section and angular distributions}

To calculate the TCS amplitude we start with the hadronic tensor
(\ref{leading-order}), evaluated in the $\gamma p$ c.m.\ with the
3-axis in the direction of $\vec{p}$. In order to preserve gauge
invariance beyond the leading-twist approximation, we use the
prescription of \cite{Guichon:1998xv} and take
\begin{eqnarray}
  \label{gauge}
T^{\alpha \beta} = 
  T^{\alpha\gamma} \Big|_{\rm eq.\,(\protect\ref{leading-order})}
  \, \left( g_{\gamma}{}^{\beta} -  q'_{\gamma}\, 
         \frac{p'^{\beta}}{p'q'}\, \right) ,
\end{eqnarray}
where the index $\beta$ refers to the virtual photon. The subtraction
term is formally suppressed by $1/Q'$ and has effects of a few percent
on the results we will present. The $\gamma p$ cross section is then
calculated from (\ref{gauge}) and the exact expression of the BH
amplitude.  We have compared the interference term thus obtained with
the approximate expressions in Sect.~\ref{sec:inter}. For $Q'^2=5
\gev^2$, $|t| =0.2 \gev^2$, and $\sqrt{s}=5 \gev$ we find that the
approximation (\ref{intres}) with (\ref{mmimi}) deviates by at most
10\% from what we obtain with (\ref{gauge}).  As one expects, the
situation gets worse for larger values of $|t|$ but improves quickly
for larger values of $Q'^2$.

%
\begin{figure}
\begin{center}
     \epsfxsize=0.52\textwidth
     \epsffile{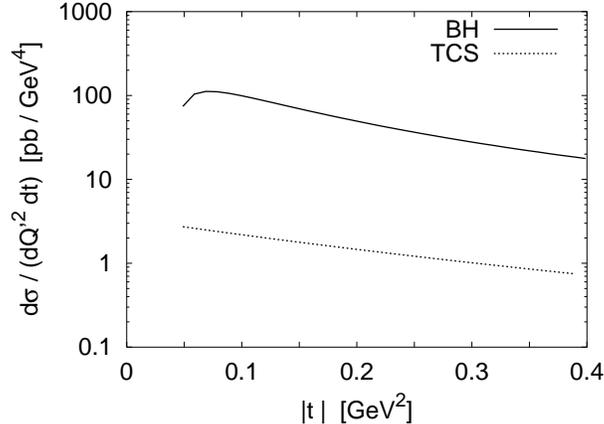}
\caption{The BH (solid line) and TCS (dotted line) cross sections for
$\sqrt{s}=5 \gev$ and $Q'^2=5 \gev^2$, integrated over $\varphi \in
[0,2 \pi]$ and $\theta \in [\pi/4,3 \pi/4]$. The TCS contribution is
calculated using $H^q_{\dd}$ and $\tilde{H}^q_{\dd}$
from~(\protect\ref{gpd-factor}), (\protect\ref{gpd-strange}).}
\label{cross4fig}
\end{center}
\end{figure}
%

We are now ready to estimate the different contributions to the cross
section. In Fig.~\ref{cross4fig} we show the result for the TCS and
the BH contributions to the $\varphi$-integrated cross section at
$\sqrt{s}=5 \gev$ and $Q'^2=5 \gev^2$. Here and in the following we
integrate over $\theta$ from $\pi /4$ to $3\pi /4$, avoiding the
region where the BH contribution becomes hopelessly large. As we
anticipated in Sect.~\ref{sec:coco} the BH process nevertheless
dominates the cross section, with TCS contributing less than 5\% in
this kinematics.  In Fig.~\ref{cross4fig} there is no contribution
from the interference between TCS and BH since the angular integration
selects charge conjugation even quantities. Let us therefore
investigate the manifestation of the interference term in the angular
distribution. We restrict ourselves to unpolarized photons here.

%
\begin{figure}
\begin{center}
     \epsfxsize=0.52\textwidth
        \epsffile{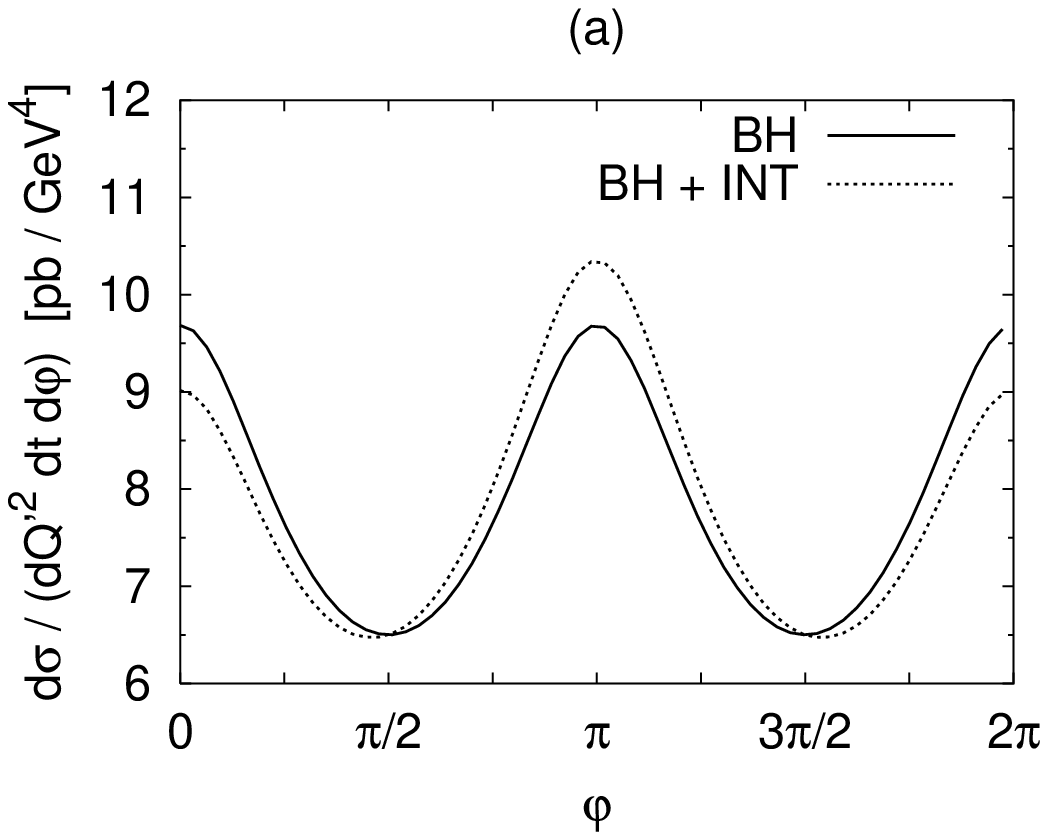}
\hspace{-0.06\textwidth}
     \epsfxsize=0.52\textwidth
        \epsffile{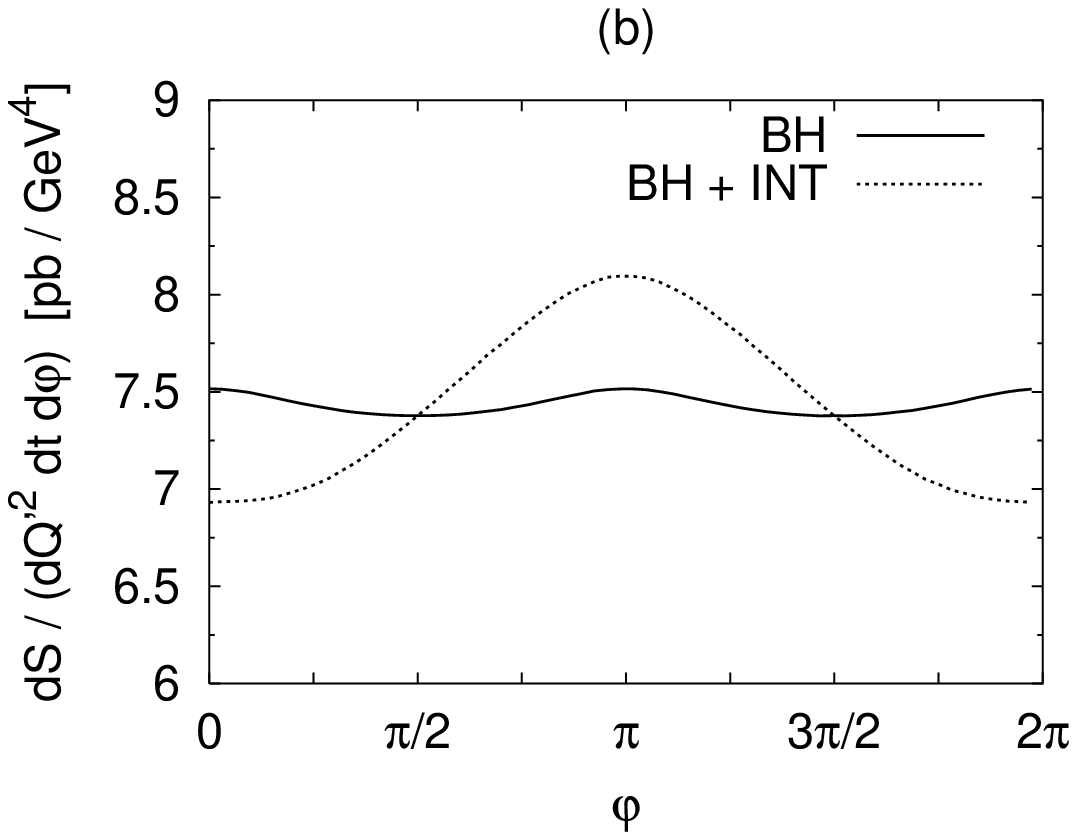}
\caption{(a) The cross section integrated over $\theta \in [\pi/4,3
\pi/4]$ as a function of $\varphi$ for $\sqrt{s}=5 \gev$, $Q'^2=5
\gev^2$, $|t| =0.2 \gev^2$.  The curves represent the BH contribution
(solid line) and the sum of BH and the interference term (dash-dotted
line), calculated using $H^q_{\dd}$ and $\tilde{H}^q_{\dd}$.  (b) The
same as in (a) but with the cross section weighted by $L/L_0$ before
integrating over $\theta$.}
\label{phi}
\end{center}
\end{figure}
%

In Fig.~\ref{phi}a we show the $\varphi$ dependence of the cross
section integrated over $\theta$ in the range $[\pi/4,3 \pi/4]$. With
the integration limits symmetric about $\theta=\pi/2$ the interference
term is odd under $\varphi\to \pi+\varphi$ due to charge conjugation,
whereas the TCS and BH cross sections are even. We separately show the
contribution from BH and the sum of BH and the interference term. The
TCS cross section is flat in $\varphi$ to leading-twist accuracy,
cf.~(\ref{TCS}), and only tiny oscillations are induced by the
prescription (\ref{gauge}). In the kinematics of the figure we get
$d\sigma_{\it TCS} /(d\qq^2\, dt\, d\varphi) \approx
0.2~\mbox{pb\,GeV}^{-4}$ when applying the same cut in $\theta$ and
taking the double distribution ansatz (\ref{gpd-factor}) for the GPDs.

Fig.~\ref{phi}b shows the corresponding contributions to the weighted
cross section
\begin{equation}
\frac{dS}{d\qq^2\, dt\, d\varphi} = 
\int_{\pi/4}^{3\pi/4}  d\theta\;
  \frac{L(\theta,\varphi)}{L_0(\theta)}\,
  \frac{d\sigma}{d\qq^2\, dt\, d\theta\,d\varphi} \, .
\end{equation}
We see that the signal is more easily visible after this
weighting. The interference term behaves now like $\cos\varphi$ up to
$1/Q'$ suppressed terms that are numerically small. The weighted BH
cross section is almost flat with our cut on $\theta$, in line with
our discussion at the end of Sect.~\ref{sec:bethe}. The TCS
contribution is again small here and will not much change the
picture. As discussed in Sect.~\ref{sec:inter}, information on the
interference can in principle also be obtained from the
$\varphi$-integrated cross section.  With the same kinematics as in
Fig.~\ref{phi} we find that the interference generates an asymmetry in
$d\sigma /(d\qq^2\, dt\, d\theta)$ about $\theta = \pi/2$ which is
barely at the 1\% level.

\begin{figure}
\begin{center}
    \epsfxsize=0.51\textwidth
    \epsffile{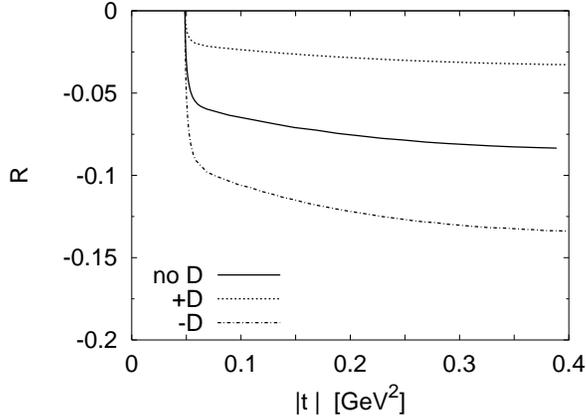}
\caption{The ratio $R$ defined in (\protect\ref{resnod}) for
$\sqrt{s}=5 \gev$ and $Q'^2=5 \gev^2$. The curves correspond to the
three models $H^q_{\protect\phantom{!}}=H^q_{\dd}$ (solid),
$H^q_{\protect\phantom{!}}=H^q_{\dd} + H^q_D$ (dotted), and
$H^q_{\protect\phantom{!}}=H^q_{\dd} - H^q_D$ (dash-dotted).}
\label{ratiofig}
\end{center}
\end{figure}

To extract information on the Compton amplitude in a compact way we
introduce
\begin{equation}
\label{resnod}
R = \frac{\displaystyle 2 \int_0^{2\pi} d\varphi\, \cos\varphi\,
          \frac{dS}{d\qq^2\, dt\, d\varphi}}{\displaystyle
\int_0^{2\pi} d\varphi\, \frac{dS}{d\qq^2\, dt\, d\varphi}} \; ,
\end{equation}
which projects out the ratio $a_1 /a_0$ of Fourier coefficients in the
weighted cross section $dS /(d\qq^2\, dt\, d\varphi) =
\sum_{n=0}^{\infty}\, a_n \cos(n\varphi)$.  Up to $1/Q'$ suppressed
contaminations the numerator in $R$ is proportional to the combination
$\tilde{M}^{--}$ of Compton amplitudes , whereas the denominator is in
our kinematics dominated by the BH part of the cross section. To
explore the dependence of our estimates on the GPDs we compare in
Fig. \ref{ratiofig} the ratio $R$ for the cases where $H^q$ is taken
from the double distribution ansatz (\ref{gpd-factor}),
(\ref{gpd-strange}) alone, or as the sum of this and the $D$-term in
(\ref{dt}). Due to a numerical accident the contributions from
$H_{\dd}$, $H_D$ and $\tilde{H}_{\dd}$ in (\ref{mmimi}) nearly cancel
each other and produce a quite small interference term. This result
should be interpreted with care since, as we discussed, $H^q_{\dd}$ is
obtained by extrapolating information from the usual parton
distributions into the ERBL region, and our $H^q_D$ is the result of a
particular dynamical model. With a generic $D$-term of the same size
one could also obtain a rather sizeable interference signal, as we see
in Fig.~\ref{ratiofig} when combining $H^q_{\dd}$ and $H^q_D$ with the
``wrong'' sign for $H^q_D$.  In accordance with our discussion at the
end of Sect.~\ref{sec:model} we conclude from this exercise that the
unpolarized interference term is highly sensitive to the behavior of
the GPDs in the ERBL region, where our modeling is least reliable.

%
%

\begin{figure}
\begin{center}
   \epsfxsize=0.51\textwidth
   \epsffile{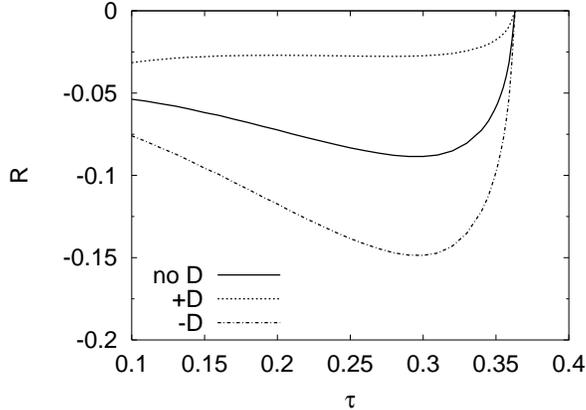}
\caption{The ratio $R$ for the same models as in
Fig.~\protect\ref{ratiofig}, but as a function of $\tau$ at $Q'^2=5
\gev^2$ and $|t| =0.2 \gev^2$.}
\label{taufig}
\end{center}
\end{figure}

Fig.~\ref{taufig} shows $R$ for the same three models of $H^q$, now as a
function of $\tau$ at fixed $Q'^2$ and $t$, and thus for varying
collision energy $\sqrt{s}$. Notice that at $\tau = 0.36$ the minimum
value of $|t|$ is equal to $0.2 \gev^2$ so that one is in collinear
kinematics, where the angle $\varphi$ is undefined. As the numerator
of $R$ projects out the coefficient of a $\cos\varphi$-dependent term
in the cross section, it must strictly vanish at that point.  We
remark that $\tau = 0.36$ is still far from its maximum value
$\tau_{\rm max}=(1 + 2 M/Q')^{-1} = 0.54$, where the production
threshold $\sqrt{s} =Q' +M$ is reached for $Q'^2 = 5 \gev^2$.  It is
interesting to note that in TCS the total collision energy at
threshold is large, whereas in DVCS or in inclusive deep inelastic
scattering one scans the resonance mass region down to the proton mass
as $x_B$ approaches its upper limit 1. While the straightforward
application of leading-twist dominance seems dangerous in TCS at
$\tau$ close to $\tau_{\rm max}$, this might be an interesting regime
to study parton-hadron duality.

%
%

We finally wish to remark on TCS with a neutron target. In that case
the BH process is suppressed due to the zero charge of the neutron. We
can explicitly see this in the approximation~(\ref{approx-BH}), where
the term in brackets involving the (typically large) factor $1/\tau^2$
goes with a combination of form factors that vanishes for $t\to 0$.
We find that the TCS contribution to the cross section is indeed more
important than for a proton target. With the kinematics in
Fig.~\ref{cross4fig} and the double distribution ansatz
(\ref{gpd-factor}), (\ref{gpd-strange}) it does however not amount to
much more than 10\% of the BH contribution. The unpolarized
interference term on the other hand generates a tiny ratio $R$ of
barely 1\%.  This can be understood from (\ref{mmimi}), where the
potentially large contribution from ${\cal H}_1$ is suppressed by the
Dirac form factor.  $\tilde{\cal H}_1$, whose prefactor survives in
the $t\to 0$ limit, is penalized with a small factor $\eta$ and
further suppressed by a near cancellation of the charge-weighted
polarized $u$ and $d$ quark densities in the neutron.

\section{Summary and discussion}
\label{sec:sum}

Next to DVCS, timelike Compton scattering may be the theoretically
cleanest process where generalized parton distributions can be
accessed.  To leading twist and at Born level, both processes involve
in fact the very same integrals over combinations of GPDs.  At the
level of $\alpha_s$ corrections and the departure from the large
$Q'^2$ limit, they will be different.  A simultaneous description of
both reactions may thus be a benchmark test of our understanding of
the dynamics, both of the approximations employed in describing the
parton level process and of the nonperturbative input.

TCS can be measured in exclusive lepton pair production, either with
quasi-real bremsstrahlung photons from incident leptons or with a
dedicated real-photon beam. Unlike DVCS, timelike Compton scattering
is always accompanied by a Bethe-Heitler contribution much bigger than
itself. It offers however relatively simple access to the real part of
the Compton amplitude via the angular distribution of the produced
lepton pair.  Appropriate angular observables allow a rather clean
investigation of the detailed structure of the Compton process.  With
circularly polarized incident photons one has access to the imaginary
parts of the Compton amplitudes and thus to the timelike analog of
what has been observed in the DVCS channel using the lepton spin
asymmetry \cite{Airapetian:2001yk}.

Using the quark handbag diagrams of Fig.~\ref{haba} and simple models
of the relevant GPDs we have estimated the cross section and angular
asymmetries for lepton pair production in a kinematical setting
typical of the HERMES regime.  We find that the angular asymmetry
carrying information on the Compton process ranges from about 5\% to
15\% within the variations of the GPD models we have explored.  This
rather wide range of predictions is generated by varying the GPDs in
the ERBL region.  It illustrates that the real part of the Compton
amplitude is highly sensitive to the form of these distributions in
the region where their physics is least known and most different from
that of ordinary parton densities.  Similar results have been obtained
in recent studies of the lepton charge asymmetry in DVCS
\cite{Kivel:2001fg,Korotkov:2001zn}.  Given the uncertainties in
modeling detailed features of GPDs, the numbers we estimate here
should hence be taken with due care. We also remark that substantial
$\alpha_s$ corrections to the DVCS amplitude have been reported for
our kinematics \cite{Belitsky:2000sg,Freund:2001rk}.  In any case,
whether the real part of TCS is observably large or not will already
provide important information about the dynamics of the Compton
process.

A look at Eq.~(\ref{mmimi}) and the plots in Figs.~\ref{hfig} and
\ref{htfig} reveals that in the region of $\tau$ we considered, the
imaginary part of the Compton amplitude is significantly larger than
its real part.  From Eqs.~(\ref{intres}) and (\ref{helasy}) it then
follows that in this kinematics the photon helicity asymmetry will be
larger with our model GPDs than the unpolarized angular asymmetry we
have investigated here.

We have not attempted to give estimates for the regime of very small
$\tau$, where DVCS has been observed at the HERA collider
\cite{Adloff:2001cn}.  In that case, the contribution from gluon GPDs
at order $\alpha_s$ is expected to be too important to be neglected.
As to the Born level quark contribution below $\tau=0.1$, we find in
our model that as $\tau$ decreases both $\re{\cal H}_1$ and $-\im{\cal
H}_1$ rise, as well as their ratio $-\re{\cal H}_1 / \im{\cal H}_1$.

Contrary to what one might expect, our estimates for a neutron target
do not give a much enhanced TCS signal in HERMES kinematics, neither
in the cross section nor in the angular distribution. This is due to
an unfortunate combination of mostly kinematic prefactors in the
formulae for an unpolarized target.  On the other hand, we do not
expect such a suppression for coherent scattering on a deuteron
target, whose GPDs have recently been discussed \cite{Berger:2001zb}.

\section*{Acknowledgments}
We gratefully acknowledge discussions with G.~Anton, R.~Baier,
J.~Bl\"umlein, J.~C.~Collins, M.~D{\"u}ren, M.~Fontannaz, A.~Freund,
T.~Gousset, B.~Kniehl, G.~Kramer, M.~McDermott, A.~Meyer, M.~Polyakov,
B.~Naroska, D.~Schiff, and G.~Sterman.  This work is supported in part
by the TMR and IHRP Programmes of the European Union, Contracts
No.~FMRX-CT98-0194 and No.~HPRN-CT-2000-00130.

\appendix
\section*{Appendix}

In this appendix we discuss to which extent the evaluation of
generalized parton distributions in the double distribution model
(\ref{ddmodel-h}) to (\ref{profile}) requires knowledge of the usual
parton distributions down to $x=0$.  This is of practical importance
since parton distributions are of course only constrained by data down
to some finite value of $x$, below which one must rely on
extrapolations.

As is well-known, the terms in (\ref{ddmodel-h}) going with
$\delta(x-x' - \eta y')$ and $\delta(x+x' - \eta y')$ individually
have non-integrable singularities at $x'=0$, but their sum is finite.
To make this explicit we consider the charge conjugation even
combination $h_+(x,\eta) = h(x,\eta) - h(-x,\eta)$, which can be
written as
%
%
\begin{eqnarray}
h_+(x,\eta) & \stackrel{x<\eta}{=} &
  \frac{1}{\eta} \int_{0}^{\frac{\eta-x}{1+\eta}} dx'\, 
  q_+(x')
   \left[ \pi\Big(x', \frac{x-x'}{\eta}\Big)
        - \pi\Big(x', \frac{x+x'}{\eta}\Big) \right]
\nonumber \\
&+& \frac{1}{\eta} 
        \int_{\frac{\eta-x}{1+\eta}}^{\frac{\eta+x}{1+\eta}} dx'\,
  q_+(x')\; \pi\Big(x', \frac{x-x'}{\eta}\Big) ,
\nonumber \\
h_+(x,\eta) & \stackrel{x>\eta}{=} &
  \frac{1}{\eta} 
        \int_{\frac{x-\eta}{1-\eta}}^{\frac{\eta+x}{1+\eta}} dx'\,
  q_+(x')\; \pi\Big(x', \frac{x-x'}{\eta}\Big) ,
\label{dd-rewrite}
\end{eqnarray}
with $q_+(x) = q(x) + \bar{q}(x)$.  Note that the corresponding
singularities at $x'=0$ are integrable for the quark valence
combination $q(x') - \bar{q}(x')$, which is not needed in the
evaluation of the Compton amplitude.  The singularities for the
polarized quark densities are integrable as well, and we can restrict
our discussion to the most problematic case (\ref{dd-rewrite}).
Writing
\begin{equation}
    \pi\Big(x', \frac{x-x'}{\eta}\Big)
  - \pi\Big(x', \frac{x+x'}{\eta}\Big)
= \frac{3}{(1-x')^3}\,
  \frac{x}{\eta}\, \frac{x'}{\eta}
\label{dd-difference}
\end{equation}
we see that the integrand in the first line of (\ref{dd-rewrite}) only
involves $x' q_+(x')$, whose singularity at $x'=0$ is integrable.

The integrals in the second and third lines of (\ref{dd-rewrite})
involve $q_+(x')$ down to values of order $x' \sim \eta-x$ and hence
are potentially problematic for $x\to\eta$.  To investigate them more
closely, we decompose
\begin{equation}
\pi\Big(x', \frac{x-x'}{\eta}\Big) = 
\frac{3}{4 (1-x'^3)} \left[\,
   \frac{x'(1-\eta) }{\eta}\, \frac{2\eta - \eta x' - x'}{\eta}
 + \frac{\eta-x}{\eta}\, \frac{\eta+x - 2x'}{\eta}  \,\right] .
\label{decompose}
\end{equation}
The first of the two terms in brackets leads again to the combination
$x' q_+(x')$ and causes no trouble when the lower integration limit
goes to zero as $x\to\eta$.  The second term does not provide a factor
$x'$ but a factor $(\eta-x)$ instead.  Let us for the sake of argument
assume that for small $x$ the quark density behaves like
\begin{equation}
  x q_+(x) \,\sim\, x^{-\lambda}
\label{small-x-behavior}
\end{equation}
with some $\lambda < 1$.  For $x\to\eta$ the integral involving the
second term in (\ref{decompose}) then goes like
\begin{equation}
|\eta - x| \int_{|\eta - x|} dx'\, q_+(x') 
   \,\stackrel{x\to\eta}\sim\,
   \frac{1}{\lambda}\, |\eta - x|^{1 - \lambda}
\label{integral-2}
\end{equation}
and hence vanishes in the limit where its evaluation requires
knowledge of $q_+(x')$ down to $x'=0$.

To get a rough feeling for the integral in the first line of
(\ref{dd-rewrite}) and for the ones involving the first term in
(\ref{decompose}) let us consider $\int dx'\, x' q_+(x')$ with lower
limit 0 and upper limit of order $\eta$.  If we assume the power
behavior (\ref{small-x-behavior}), then the contribution from the
interval $x' \in [0,\epsilon]$ to the total integral is of order
$(\epsilon /\eta)^{1-\lambda}$. For typical values of $\lambda$ this
is about 10\% if $\epsilon$ is one to two orders of magnitude smaller
than $\eta$. Of course $q_+(x)$ is unknown below some value of $x$,
but unless its small-$x$ behavior is much steeper than
(\ref{small-x-behavior}), the above estimate should not be altered
significantly.

Our discussion can be adapted to profile functions $\pi(x',y')$ other
than the one in (\ref{profile}).  Provided that $\pi(x',y')$ is
differentiable in $y'$ one can replace (\ref{dd-difference}) with
\begin{equation}
    \pi\Big(x', \frac{x-x'}{\eta}\Big)
  - \pi\Big(x', \frac{x+x'}{\eta}\Big) = 
           - \frac{2x'}{\eta}\, 
             \partial_2\pi\Big(x',\frac{x}{\eta}\Big) + O(x'^2) .
\end{equation}
where $\partial_2\pi = \partial\pi(x',y')/ \partial y'$. Decomposing
\begin{equation}
\frac{x-x'}{\eta} = (1-x') - \frac{x'(1-\eta)}{\eta}
                           - \frac{\eta-x}{\eta} 
\end{equation}
we further see that (\ref{decompose}) can be replaced with
\begin{eqnarray}
\pi\Big(x', \frac{x-x'}{\eta}\Big) &=& 
    {}- \frac{x'(1-\eta)}{\eta}\, \partial_2 \pi(x',1-x') 
  + O(x'^2) 
\nonumber \\
 && {}- \frac{\eta-x}{\eta}\, 
        \partial_2 \pi\Big(x',1-x'-\frac{x'(1-\eta)}{\eta}\, \Big) 
  + O\Big((\eta-x)^2\Big) ,
\label{dec-gen}
\end{eqnarray}
provided again that $\pi(x',y')$ is differentiable in $y'$ and that in
addition $\pi(x',1-x')=0$.

Notice that while the GPD so obtained is finite at $x=\eta$, its first
derivative in $x$ is in general not.  For profile functions
$\pi(x',y')$ vanishing at $x'+y'=1$ we readily obtain a representation
of $\partial h_+ /\partial x$ analogous to (\ref{dd-rewrite}) with
$\pi$ replaced by $\eta^{-1} \partial_2 \pi$.  Provided that
$\partial_2\pi$ is differentiable in $y'$ our previous line of arguments
goes through, except for the equivalents of (\ref{decompose}) and
(\ref{dec-gen}).  There one will have an extra term $\eta^{-1}
\partial_2 \pi(x',1-x')$ on the right-hand side, which for our profile
function (\ref{profile}) is nonzero.  This means that the integrals
corresponding to the second and third lines of (\ref{dd-rewrite}) will
have an integrand going like $q_+(x')$ at $x'\to 0$, without a factor
$(\eta-x)$ in front.  $\partial h_+ /\partial x$ thus behaves for
$x\to \eta$ like $\int dx'\, q_+(x')$ with lower limit of order
$|\eta-x|$, and with the small-$x$ behavior (\ref{small-x-behavior})
diverges like
\begin{equation}
\left| \frac{\partial}{\partial x}\, h_+(x,\eta) \,\right|
  \,\stackrel{x\to\eta}{\sim}\, 
  \frac{1}{\lambda}\, |\eta - x|^{- \lambda} .
\label{derivative}
\end{equation}
Now, the principal value integral over $x$ which gives $\re {\cal
H}_1$ according to (\ref{htilde}) effectively involves $\partial h_+
/\partial x$ at $x$ around $\eta$, but the singularity
(\ref{derivative}) is integrable and gives a finite result for the
amplitude.  One can also insert the expression (\ref{ddmodel-h}) into
(\ref{htilde}) and explicitly carry out the integrals over $x$ and
$y'$ for the profile function (\ref{profile}). The result for $\re
{\cal H}_1$ has the form $\int_0^1 dx'\, x' q_+(x')\, \rho(x',\eta)$,
where $\rho(x',\eta)$ has a $\log(x')$ singularity at $x'=0$. Up to
this logarithm the small-$x'$ behavior of the quark density thus
enters $\re {\cal H}_1$ in the same way as according to our above
discussion it enters $h_+(\eta,\eta)$ and hence $\im {\cal H}_1$.

\begin{figure}
\begin{center}
   \epsfxsize=0.49\textwidth
   \epsffile{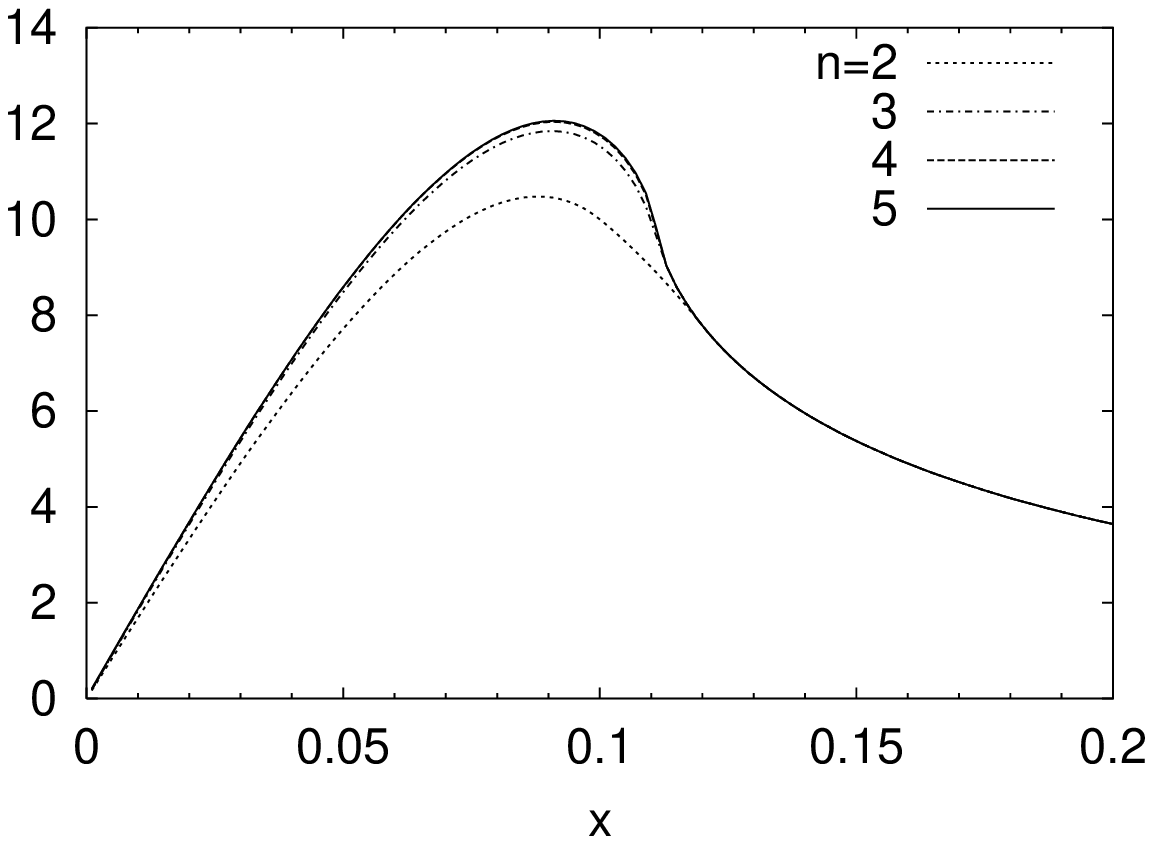}
\caption{$h^u_+(x,\eta)$ for $\eta=0.11$, corresponding to $\tau=0.2$,
evaluated with a lower cutoff $\epsilon = 10^{-n}$ on $x'$ in the
integrals (\protect\ref{dd-rewrite}).}
\label{fig:dd-cut}
\end{center}
\begin{center}
   \epsfxsize=0.49\textwidth
     \epsffile{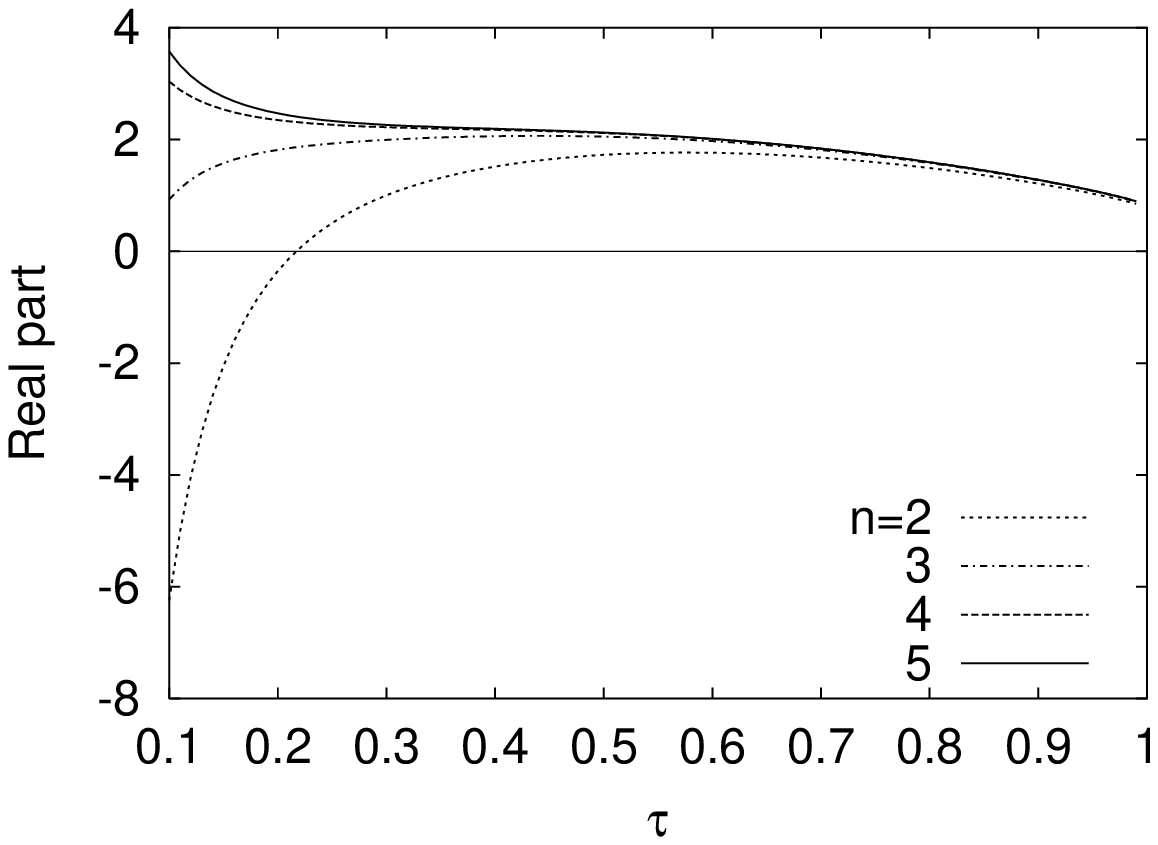}
   \epsfxsize=0.49\textwidth
     \epsffile{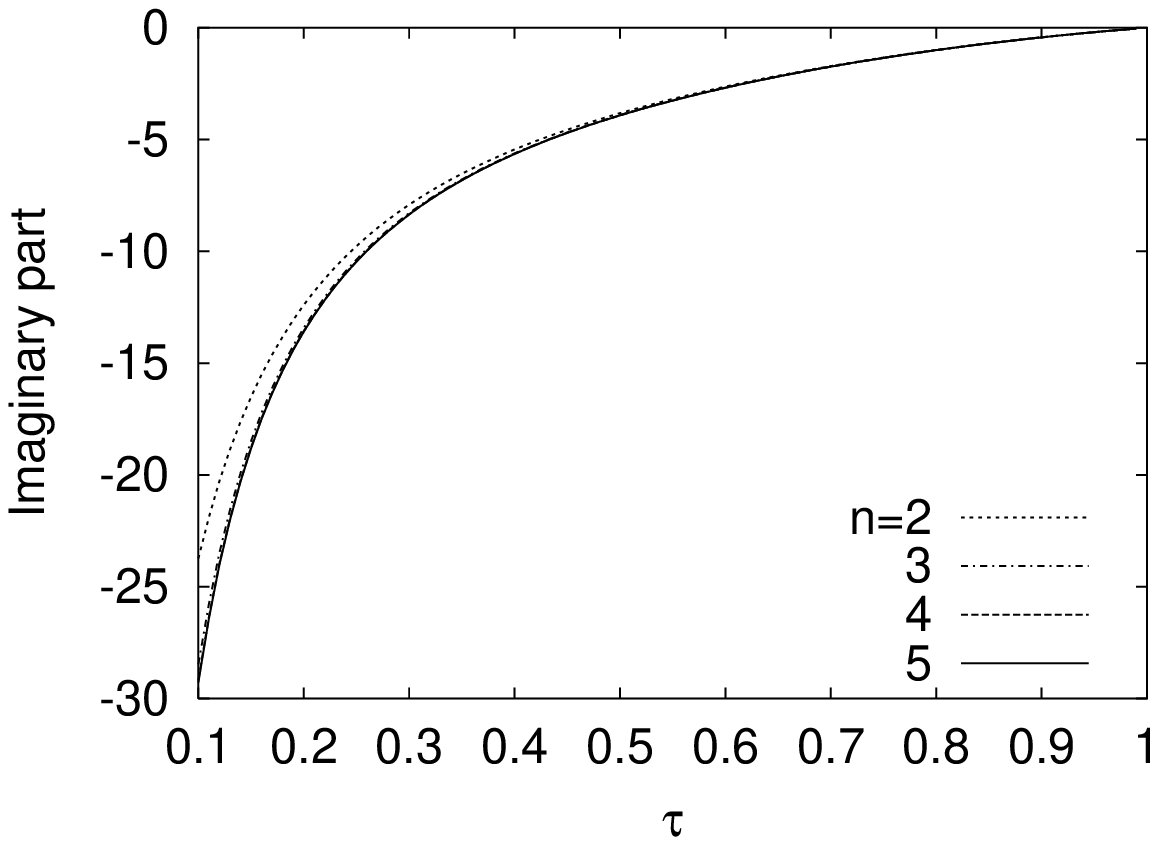}
\caption{$\re{\cal H}_1^u$ (left) and $\im{\cal H}_1^u$ (right)
divided by $\frac{1}{2} F^u_1(t)$, calculated from the GPDs in
Fig.~\protect\ref{fig:dd-cut}.}
\label{fig:H-cut}
\end{center}
\end{figure}

Let us explore how our arguments work at the quantitative level,
restricting ourselves to $u$ quarks for simplicity.  In
Fig.~\ref{fig:dd-cut} we plot $h^u_+(x,\eta)$ at $\eta=0.11$ as it is
obtained from (\ref{dd-rewrite}) when setting $u_+(x')=0$ for $x'$
below some cutoff $\epsilon$.  We find good convergence as $\epsilon$
approaches zero, the curves for $\epsilon=10^{-4}$ and $10^{-5}$ being
hardly distinguishable.  All predictions in this paper have been
obtained with $\epsilon= 10^{-5}$.  Fig.~\ref{fig:H-cut} shows
$\re{\cal H}_1^u$ and $\im{\cal H}_1^u$ as functions of $\tau$,
calculated with the same cutoffs as in Fig.~\ref{fig:dd-cut} when
constructing the GPDs.  Clearly the convergence is much slower for the
real part, which can be traced back to large cancellations in the
relevant integrals.\footnote{For the numerical evaluation of
(\protect\ref{htilde}) we add and subtract $H(x,\eta,t)$ at the points
$|x|=\eta$.  To avoid cancellations as much as possible we only do
this for $|x|< 2\eta$ when $\eta< 1/2$, writing
\begin{eqnarray*}
\mathrm{PV}\! \int_0^1 dx\, 
        \left[ \frac{1}{x-\eta} - \frac{1}{x+\eta} \right] h_+(x,\eta)
&\!=\!& \int_0^{2\eta} dx\, 
        \left[ \frac{1}{x-\eta} - \frac{1}{x+\eta} \right] 
         \Big[ h_+(x,\eta) - h_+(\eta,\eta) \Big]
+ h_+(\eta,\eta)\, \ln\frac{1}{3}
\nonumber \\
&& \hspace{-0.8em} {}+ \int_{2\eta}^1 dx\, 
        \left[ \frac{1}{x-\eta} - \frac{1}{x+\eta} \right] 
         h_+(x,\eta) ,
\end{eqnarray*}
where PV denotes the principal value prescription.}
This illustrates again the sensitivity of $\re{\cal H}_1$ to small
changes of the GPDs we have already encountered in
Sect.~\ref{sec:model}.

\begin{figure}
\begin{center}
   \epsfxsize=0.49\textwidth
     \epsffile{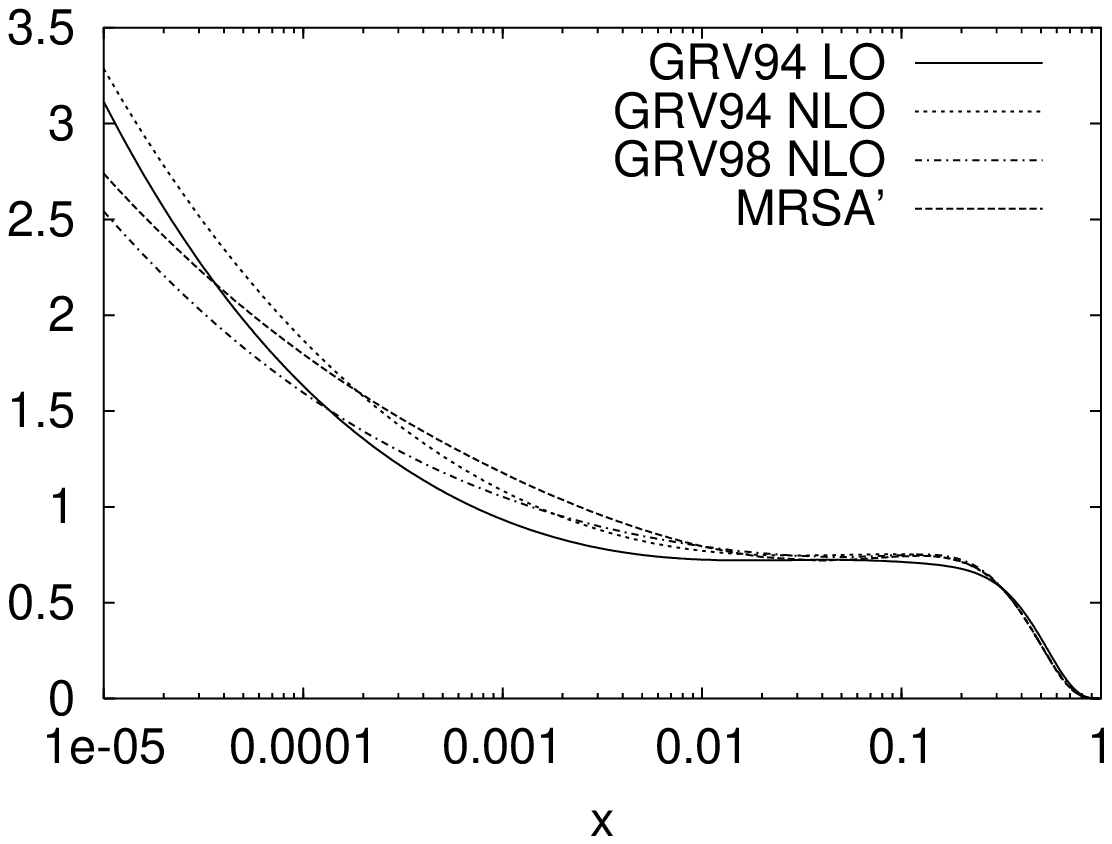}
   \epsfxsize=0.49\textwidth
     \epsffile{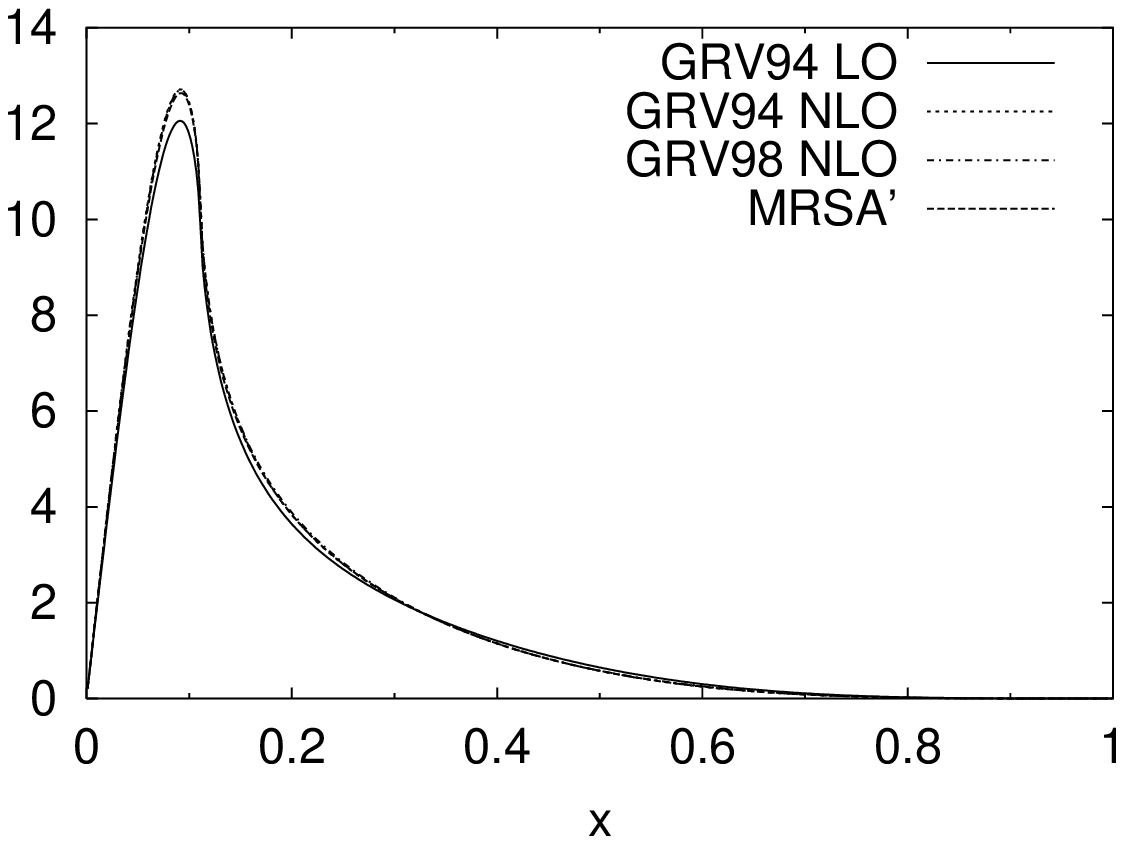}
\caption{Left: Different parameterizations of $u(x)+\bar{u}(x)$ for
$\mu^2= 5 \gev^2$.  Right: The corresponding generalized distributions
$h_+^u(x,\eta)$ at $\eta=0.11$, obtained with our double distribution
ansatz.}
\label{fig:par}
\end{center}
\begin{center}
   \epsfxsize=0.49\textwidth
     \epsffile{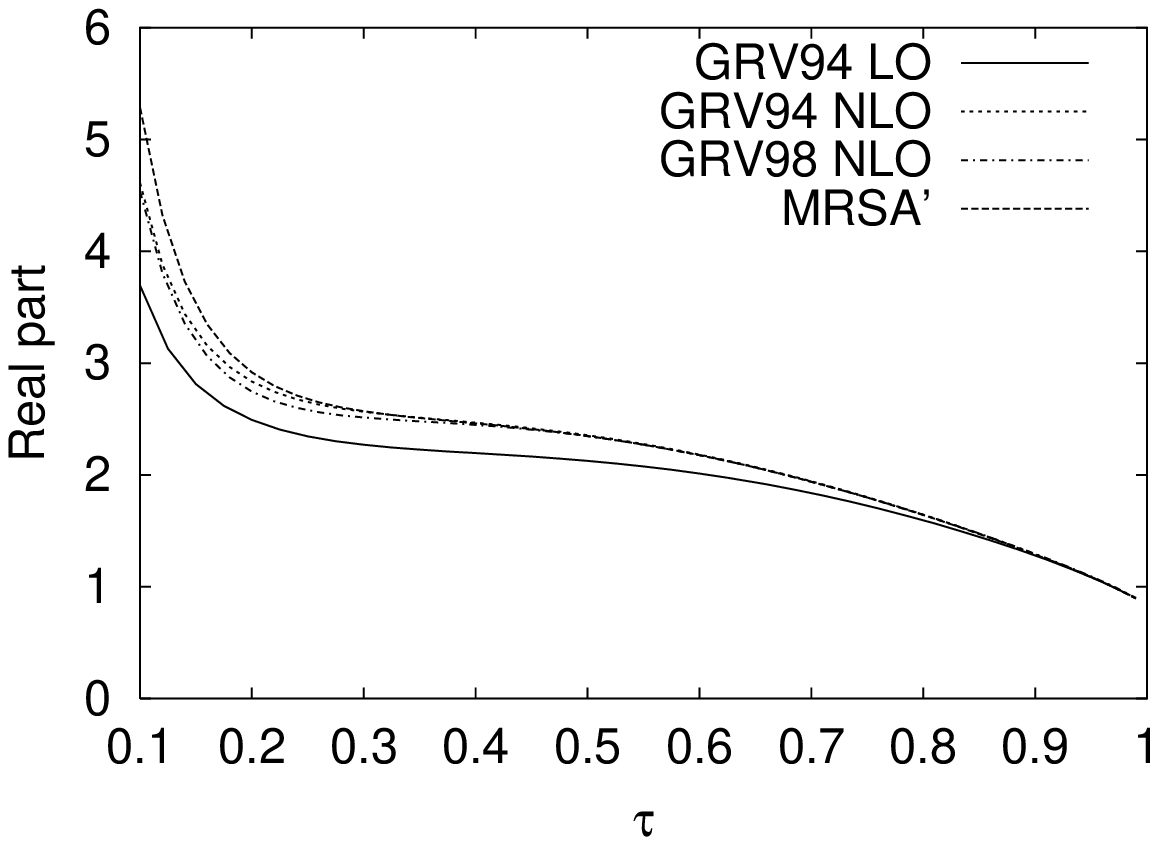}
   \epsfxsize=0.49\textwidth
     \epsffile{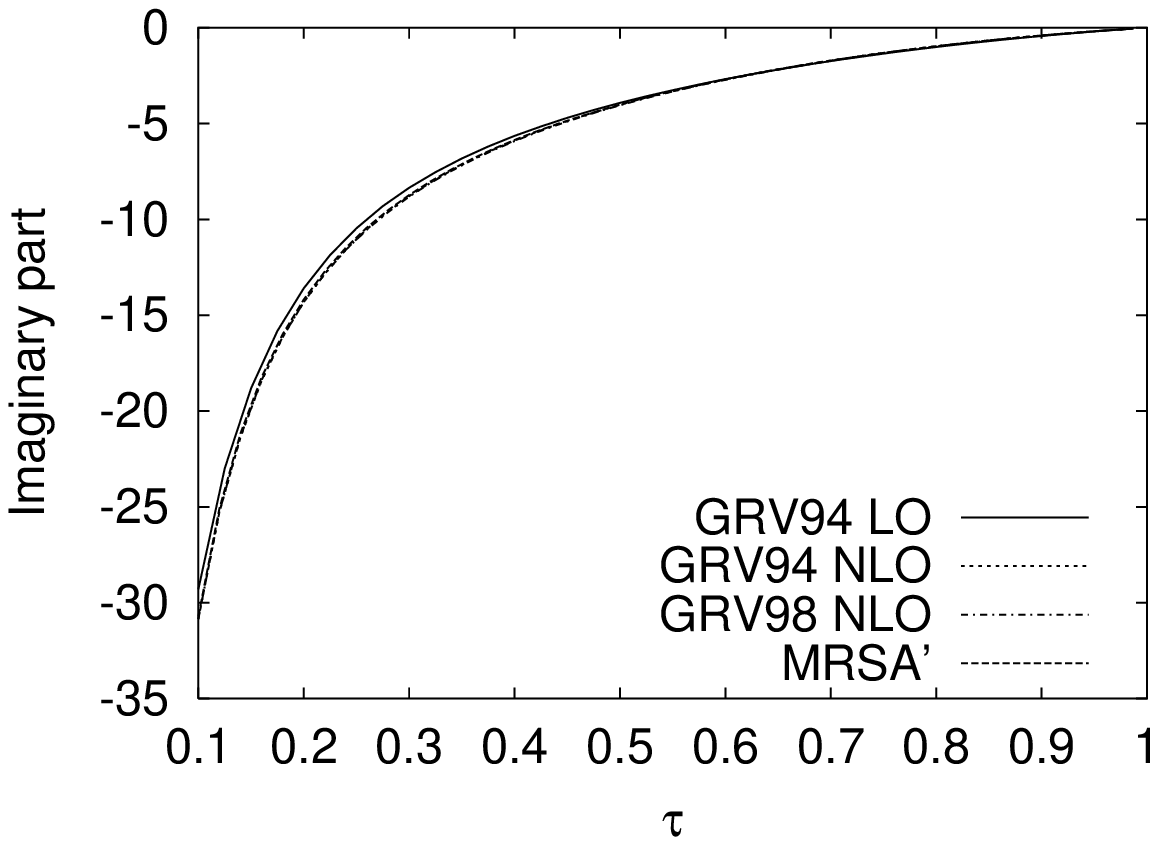}
\caption{$\re{\cal H}_1^u$ (left) and $\im{\cal H}_1^u$ (right)
divided by $\frac{1}{2} F^u_1(t)$, calculated from the same
parameterizations of $u(x)+\bar{u}(x)$ as in
Fig.~\protect\ref{fig:par}.}
\label{fig:par-H}
\end{center}
\end{figure}

In Fig.~\ref{fig:par} we show $h_+^u(x,\eta)$ obtained from different
parameterizations of the usual quark densities at $\mu^2= 5 \gev^2$,
calculated with a lower cutoff $\epsilon= 10^{-5}$ on $x'$.  We
compare the GRV 94 LO parameterization used in our predictions with
three NLO distributions in the $\mathrm{\overline{MS}}$ scheme: GRV 94
NLO \cite{Gluck:1995uf}, GRV 98 NLO \cite{Gluck:1998xa}, and MRSA$'$
\cite{Martin:1995ws}.  All input densities are clearly distinct for
$x$ below $0.01$, but above $x\approx 0.05$ the three NLO
parameterizations hardly differ among themselves.  The corresponding
curves for $h_+^u(x,\eta)$ at $\eta=0.11$ are almost identical for the
NLO parameterizations, in accordance with our arguments about the
relevance of small $x$ in the input densities.  In
Fig.~\ref{fig:par-H} we make the same comparison for $\re{\cal H}_1^u$
and $\im{\cal H}_1^u$.  In line with our previous findings,
differences between the parameterizations are more prominent for
$\re{\cal H}_1^u$, but they remain quite small.  We thus do not
confirm the results of \cite{Freund:2001rk,Freund:2001bf}, where
substantially different GPDs and Compton amplitudes have been obtained
from the GRV 98 NLO and MRSA$'$ distributions with the same ansatz
(\ref{ddmodel-h}), (\ref{profile}) we have used here.\footnote{Our
results do not change qualitatively if we implement the double
distribution ansatz (\protect\ref{ddmodel-h}) not at $\mu^2=5 \gev^2$
but at $\mu^2=4 \gev^2$ as was done in
\protect\cite{Freund:2001rk,Freund:2001bf}. The origin of our
discrepancies with \protect\cite{Freund:2001rk,Freund:2001bf} is under
investigation with the authors.}

In summary, we have discussed in detail the calculation of GPDs with
the double distribution ansatz (\ref{ddmodel-h}). In the DGLAP region
the calculation of $h_+(x,\eta)$ involves the usual quark densities
down to $x' = (x-\eta) /(1-\eta)$, whereas in the ERBL region the
ansatz literally does require knowledge of $q_+(x')$ down to $x'=0$.
The corresponding integrals involve however only the combination $x'
q_+(x')$, or they vanish like (\ref{integral-2}).  The contribution
from $q_+(x')$ at values of $x'$ several orders of magnitude below
$\eta$ should thus be moderate, unless one has an extremely steep rise
of the quark density at small $x'$.  We confirm this numerically by
evaluating the integrals (\ref{dd-rewrite}) with different lower
cutoffs on $x'$.  For $\eta$ of order 0.1 we find rather similar GPDs
when implementing the ansatz (\ref{ddmodel-h}) with different
parameterizations of the quark densities, given that they show only
mild discrepancies down to $x'$ of order $10^{-2}$.

The derivative of $h_+(x,\eta)$ at $x\to\eta$ becomes infinite with
the profile function we used in our double distribution ansatz, but
the corresponding singularity in $x$ is integrable.  The small-$x$
behavior of $q_+(x)$ is found to affect the real and imaginary parts
of the Born level Compton amplitude in a similar way, with the real
part showing somewhat higher sensitivity.  Again we confirmed this in
our numerical study.

We finally emphasize that our discussion of how relevant the usual
quark densities at very small $x$ are in the construction of GPDs
refers to a particular \emph{model} prescription.  It is a different
question to what extent one \emph{physically} expects the behavior of
$q_+(x)$ at $x\to 0$ to be reflected in GPDs at finite $\eta$.  In the
representation of GPDs as the overlap of wave functions for the
incoming and outgoing hadron \cite{Brodsky:2001xy} one finds indeed
that for $x\to\eta$ a quark momentum in \emph{one} of the two wave
functions goes to zero.  This is similar to, but not the same as the
situation for a usual quark density at $x\to 0$, where \emph{both}
wave functions involve a quark with zero momentum.


\end{document}